# Imaging Electronic Correlations in Twisted Bilayer Graphene near the Magic Angle

Youngjoon Choi[1,2,4], Jeannette Kemmer[1,2], Yang Peng [2,3,4], Alex Thomson[2,3,4], Harpreet Arora[1,2], Robert Polski[1,2], Yiran Zhang[1,2,4], Hechen Ren[1,2], Jason Alicea[2,3,4], Gil Refael[2,3,4], Felix von Oppen[2,5], Kenji Watanabe[6], Takashi Taniguchi[6], and Stevan Nadj-Perge[1,2*]

[1]T. J. Watson Laboratory of Applied Physics, California Institute of Technology, 1200 East California Boulevard, Pasadena, California 91125, USA

[2]Institute for Quantum Information and Matter, California Institute of Technology, Pasadena, California 91125, USA

[3]Walter Burke Institute for Theoretical Physics, California Institute of Technology, Pasadena, California 91125, USA

[4]Department of Physics, California Institute of Technology, Pasadena, California 91125, USA

[5]Dahlem Center for Complex Quantum Systems and Fachbereich Physik, Freie Universität Berlin, 14195 Berlin, Germany

[6]National Institute for Materials Science, Namiki 1-1, Tsukuba, Ibaraki 305 0044, Japan

**Abstract:** Twisted bilayer graphene with a twist angle of around 1.1° features a pair of isolated flat electronic bands and forms a strongly correlated electronic platform. Here, we use scanning tunneling microscopy to probe local properties of highly tunable twisted bilayer graphene devices and show that the flat bands strongly deform when aligned with the Fermi level. At half filling of the bands, we observe the development of gaps originating from correlated insulating states. Near charge neutrality, we find a previously unidentified correlated regime featuring a substantially enhanced flat band splitting that we describe within a microscopic model predicting a strong tendency towards nematic ordering. Our results provide insights into symmetry breaking correlation effects and highlight the importance of electronic interactions for all filling factors in twisted bilayer graphene.

*Corresponding author: s.nadj-perge@caltech.edu

Electronic properties of metals, insulators, and semiconductors are frequently described in the framework of non-interacting electrons. This description works remarkably well when the kinetic energy of electrons at the Fermi energy, as set by the dispersion of the electronic bands, is large compared to the Coulomb interaction. However, the kinetic energy is quenched in materials with almost non-dispersive (flat) bands. In these materials, interactions between electrons may give rise to the formation of novel strongly correlated electronic phases. Twisted bilayer graphene (TBG) has recently emerged as a highly tunable and surprisingly simple platform for the experimental investigation of such strongly correlated phases (*1*–*3*). Two layers of graphene are stacked with a twist angle of $\theta$ and form a periodic moiré pattern (Fig. 1A), localizing electronic wave functions (*4*–*7*) in regions in which carbon atoms are stacked on top of each other (AA stacking).

While monolayer graphene is well described in terms of free Dirac electrons (*8*), in a bilayer hybridization significantly alters the electronic band structure – particularly when the layers are twisted by a small angle $\theta < 3^{\circ}$ (*9*–*17*). In this case, interlayer hybridization results in the formation of two low energy bands separated by a gap from the higher energy dispersive bands (Fig. 1B and 1C) (*5*–*7*, *18*–*20*). The resulting van Hove singularities (vHs) generate maxima in the local density of states and can be probed directly using scanning tunneling microscopy (STM) (*10*). As the twist angle is reduced, the vHs move closer together. Finally, close to the magic angle value $\theta \approx 1.1^{\circ}$ (*5*) the bands become flat.

Near the magic angle, electron-electron interactions dramatically modify the electronic properties of TBG due to the quenching of kinetic energy. Interaction effects have been observed in recent transport experiments (*1*–*3*, *17*), which reveal the existence of correlated insulating and superconducting states. The strong correlations along with the large number of atoms in a moiré unit cell pose substantial challenges to a microscopic understanding of this system. Local probes such as STM can thus provide invaluable information, both on the electronic band structure and correlation effects.

Figure 1D shows a sketch of the experimental setup. The starting point is a van der Waals structure consisting of twisted bilayer graphene placed on top of a thin boron nitride (BN) layer. An approximately 10 nm thick metallic graphite multilayer is placed underneath the BN and serves as a back gate. The twisted bilayer and the back gate are contacted separately using predefined gold electrodes. The main difference with previous STM measurements (*14*–*16*) on gated bilayer graphene is our careful alignment of the graphene layers close to the magic angle and the use of a graphite back gate that is shown to reduce disorder effects (*3*, *21*). Unlike the samples used for transport studies, the TBG in our experiments is not covered by BN; see Fig. 1E and (*22*) section 1. In line with previous experiments (*3*, *14*, *23*), we find that our TBG is not uniform across the entire sample, forming domains with different local twist angles. We show results from two devices with domains exhibiting angles in the range $1^{\circ} < \theta < 2^{\circ}$ measured at an effective temperature of T=1.5K as calibrated on the superconducting gap of a lead crystal.

Figures 1F and 1G show topographies of two devices with distinct moiré superlattice periodicities of $L_M \approx$ 7nm (Fig. 1F, device D1) and $L_M \approx$ 13nm (Fig. 1G, device D2). We estimate a local twist angle θ from the area of the moiré unit cell and find $\theta = 1.92^{\circ}$ and $\theta = 1.01^{\circ}$, respectively. The topographies show an apparent periodic height increase of approximately 3-8Å which reflects the increased local density of states of the AA stacking regions, confirming the formation of a moiré lattice (*6*). We typically observe that the moiré superlattice is not strictly threefold symmetric as the periods in the three directions can differ by approximately 5-10%.

Figures 1H and 1I show the observed point spectra for the non-magic twist angle θ=1.92˚ at various back gate voltages $V_{Bg}$. All point spectra are taken on AA regions where the flat bands predominately localize. The two peaks in the local tunneling density of the states (TDOS) are identified as vHs separated by approximately 150meV. This splitting is consistent with previous STM results for similar angles (*10*). Changing the position of the Fermi level relative to the vHs via the back gate shifts the positions of the two peaks in bias voltage, but leaves their separation and overall line shape approximately unchanged. This is consistent with a simple band-structure picture (see (*22*), section 2, for data at larger $\theta$).

For regions near the magic angle ($\theta$ =1.01˚), the observed spectra differ qualitatively (Figs. 1J and 1K). Here, the overall peak shapes change dramatically as they approach the Fermi energy ($V_{Bias}$ = 0mV). The peak approaching the Fermi level sharpens and increases in height, signaling a rapid decrease in lifetime broadening. This feature is seen for both electron and hole doping. Moreover, the splitting between the peaks is highly sensitive to $V_{Bg}$ (see also Fig. 2 and the corresponding discussion) indicating that electronic correlations cause deviations from a simple band-structure picture of the flat bands.

The detailed evolution of the local TDOS as a function of electron density for device D1 (θ=1.04˚) is shown in Fig. 2A. Higher doping levels could be accessed in device D1 due to a slightly thinner gate dielectric. When the electronic flat bands sit below the Fermi level (3.8V< $V_{Bg}$ <10V), the completely filled bands shift linearly as $V_{Bg}$ is reduced. The slope of $\Delta V_{Bg}/\Delta V_{Bias} \approx$ 1V/5mV is directly proportional to the density of states (DOS) Δn/Δμ at the Fermi level in the regimes where electronic correlations are weak (with n being the carrier concentration and μ the chemical potential). When the DOS is high, the slope $\Delta V_{Bg}/\Delta V_{Bias}$ is large as substantial changes in back-gate voltage are needed to shift the chemical potential. In the opposite limit, when the Fermi level passes through a gapped region, the slope $\Delta V_{Bg}/\Delta V_{Bias}$ is close to zero. The linear shift of the peaks for 3.8V< $V_{Bg}$ < 10V indicates a constant DOS as expected when the Fermi level is located in the non-flat (dispersive) bands, as illustrated in Fig. 2B.

The slope is reduced in the gate range 2.7V< $V_{Bg}$ < 3.8V (Fig. 2A, orange line) just before the flat bands start to cross the Fermi level. This observation indicates a gap between the flat and the upper dispersive band, of approximately 25meV according to the total shift in the chemical potential over this range of $V_{Bg}$ . The finite slope in this region signals a remnant density of subgap states that likely originates from localized states often observed in STM measurements (*9*, *10*, *16*). We also find that the gap decreases for smaller twist angles (Fig. 2C) as predicted theoretically (*24*). The extracted gap values strongly suggest that the additional bands become increasingly important as the angle drops below 1° (*5*), thus setting a lower bound on the twist angle for observing correlated insulating states.

For -9V< $V_{Bg}$ <2.7V, the bands become significantly distorted due to electronic correlations. We observe several suppressions of the TDOS at the Fermi level, and the slope $\Delta V_{Bg}/\Delta V_{Bias}$ changes repeatedly. First, the slope is large as the upper flat band with its large DOS crosses the Fermi level (-1V< $V_{Bg}$ < 3V). In this region, suppressions of the TDOS near $V_{Bias}$ = 0mV are observed, as discussed in more detail in Fig. 4. As the upper flat band is depopulated, the slope $\Delta V_{Bg}/\Delta V_{Bias}$ of the upper vHs peak sharply decreases, while the slope of the lower peak barely changes, enhancing the apparent splitting between the bands. At $V_{Bg} \approx$ -3V, the splitting becomes maximal. Changing $V_{Bg}$ further, the splitting of the peaks diminishes (for -6V< $V_{Bg}$ <-3V) and then peak corresponding to the lower flat band passes through the Fermi level exhibiting several TDOS suppressions (for -9V< $V_{Bg}$ <-6V). The total range of $\Delta V_{Bg} \approx$ 13V over which the two flat bands cross the Fermi level is expected to change the charge density by $\Delta q = 8e^-/A$ = 1.6e-19C×5.46e12 $cm^{-2}$ (A is the measured area of this moiré unit cell) corresponding to filling of the two bands by a total of 8 electrons. This matches the change in charge density extracted experimentally based on the gate capacitance for a BN thickness $d_{BN}$ = 40nm as measured by AFM, and a dielectric constant $\varepsilon_{BN} \approx 3$. The relation between the $V_{Bg}$ range and a density change of $8e^-/A$ is accurate to within 7-10% and allows us to reliably determine $V_{Bg}$ values corresponding to fractional fillings of the flat bands.

One of the surprising results of the experiment is the enhanced splitting between TDOS peaks around the charge neutrality point (CNP) compared to the splitting of the fully occupied bands $\Delta_{CNP}$ = 15±5meV, Figs. 2D and 2E. This enhancement only stands out in a full bias-versus-back gate density map and hence has not been reported in transport measurements, which are only sensitive to energy scales of a few meV from the Fermi level. Exchange interactions are plausible origin of the enhanced splitting at the CNP in analogy with quantum Hall ferromagnetism (*25*, *26*). In this scenario, the strong exchange interaction pushes the bands apart when one of the flat bands is completely filled and the other completely empty. Two important differences distinguish TBG from quantum Hall systems. First, no external field is needed as flat bands are already present at zero field; and second, the exchange interaction may not open a hard gap. Certain points in the Brillouin zone (away from the flat regions of the band causing the vHs) may remain gapless as a result of the preserved symmetries (*27*, *28*).

We support this interpretation by calculations within the framework of a ten-band model for TBG developed in Ref. (*29*). The model captures the non-interacting band structure of magic-angle TBG and incorporates all relevant symmetries. Although its overall bandwidth for the flat bands and the vHs peak splitting (7meV and 1mV, respectively) are small compared to our observations, the model provides qualitative guidance on interaction effects, including broken symmetries. To account for interaction effects, we have added a symmetry-preserving short-range interaction of strength $E_c$ (see (*22*), section 5, for details) and solved the model self-consistently within mean-field theory assuming that the symmetry between the four flavors (spin and valley) remains unbroken. We find several nearly degenerate broken-symmetry states (see (*22*), section 5, for additional discussion). The solution breaking $C_3$ symmetry qualitatively reproduces the observed splitting near the CNP (Fig. 3). Other features such as the overall broadening of the bands near CNP are also qualitatively captured. The resulting solution shows that the peaks in the LDOS are split by an amount set by $E_c$. For $1< E_c/W <2$, with W being the bandwidth of the non-interacting model, the relative increase in the splitting matches well with our experimental observations (Fig. 3A-C). The competitive energy of the $C_3$-breaking solution suggests a high susceptibility towards nematicity (*30*, *31*); effects such as sample strain and unaccounted electronic correlations can plausibly stabilize this order as the unique ground state. Figures 3D and 3E show the corresponding band structure across the Brillouin zone indicating the presence of two Dirac points protected by $C_2\mathcal{T}$ symmetry. Moreover, the model predicts that the breaking of the $C_3$ symmetry is spatially most pronounced in the LDOS at the bridges connecting different AA sites for energies close to vHs peaks, as calculated in Fig. 3F.

Consistent with these theoretical findings, spatially resolved TDOS maps taken on sites within the most $C_3$ symmetric moiré regions show strongly anisotropic TDOS profiles near the CNP (Fig. 3G). Experimentally, the most pronounced anisotropy is observed near the Fermi level when the TDOS at AA sites is suppressed, thus highlighting the bridge areas. Small energy-dependent variations for different directions are however noticeable even when the TDOS at AA sites is larger, signaling competition between the effects of strain and electronic correlations. Notably, the most pronounced bridge direction changes dependent on whether $V_{Bias}$, is adjusted to probe the lower or upper flat-band peak as expected from the model calculations (Fig. 3F). We note that the moiré periodicities in this particular area are $L_{MA}$ = 13.8 nm; $L_{MB}$ = 13.7nm; and $L_{MC}$ = 14.2nm, indicating the presence of only 0.1%-0.2% of uniaxial heterostrain (*32*,

*33*). These values of external strain correspond to energy scales of 0.1-0.5meV which is much smaller than the energies corresponding to TBG lattice relaxation effects (*24*) or electronic correlations measured here. This implies that strain alone is unlikely to explain the measured anisotropies and suggests the formation of a nematic ground state of TBG near the CNP.

Finally, we focus on the states near half filling of the flat bands (Fig. 4). Figure 4A shows the point spectroscopy as a function of $V_{Bg}$ for a moiré site with a local twist angle θ = 1.01° (D2). In addition to the interaction-enhanced splitting at the CNP, a suppression in TDOS is observed near half filling both for electrons and holes, signaling the emergence of correlated gaps. Line cuts close to half filling (Fig. 4B and 4C) show gap values between 2 and 4 meV that are slightly higher than the values extracted from thermal activation in transport experiments (*1*, *3*). We suspect that this difference reflects effective disorder averaging over the sample area in transport measurements which naturally reveals a reduced gap.

To better resolve the dependence on the flat-band filling, we plot the zero-bias TDOS as a function of $V_{Bg}$ in Fig. 4D. The trace shows local minima of the zero-bias TDOS for half filling of the flat bands, consistent with correlated ground states at these fillings. The observed gaps typically occur near half filling, although they are sometimes offset, presumably due to local electrostatic disorder, strain, or tip-related effects. Occasionally the less developed states at one-quarter and three-quarter filling may also be resolved (black arrows in Fig. 4D). These observations strongly indicate that the measured gaps originate from correlated states at commensurate fillings and are distinct from other effects such as the Coulomb gap (*34*) that can also suppress the TDOS near the Fermi level in 2D systems with strong electronic correlations (*35*–*40*).

Our results provide a local, spatially resolved picture of the electronic phases of TBG near the magic angle. We found that the band structure of the flat bands is considerably broader than anticipated and comparable to the Coulomb interaction energy scale. At half filling, we observed gaps for both electrons and holes which are consistent with the emergence of correlated insulating states. Our results show that these interaction effects are robust against small deviations (of the order of 0.1˚) from the magic angle. This suggests that strain and disorder play more important roles for the observability of correlated states than an extremely precise angle alignment. For twist angles below 0.9˚-1˚, however, the band gap between the dispersive and flat bands becomes considerably smaller (*24*). This sets a lower bound on the twist-angle range for which correlated flat band physics can be accessed, as the additional bands become relevant for smaller angles (*5*, *24*). At charge neutrality, our results indicate that the ground state exhibits nematic order and breaks $C_3$ symmetry due to exchange interactions. Theoretically, one would expect that a similar exchange-driven mechanism opens a hard gap at fractional fillings when spin and valley degrees of freedom are taken into account (*25*, *41*). How superconductivity emerges from such symmetry broken correlated insulating states remains an open question.

Note added: During preparation of this manuscript, we become aware of related work (arXiv:1812.08776).

**Acknowledgments**: We gratefully acknowledge discussions with Ray C. Ashoori, Pablo Jarillo-Herrero, Ashvin Vishwanath, James Eisenstein, Andrea Young, and Haim Beidenkopf. **Funding:** The STM work is in part supported by NSF through program CAREER DMR-1753306. Sample fabrication efforts are supported by DMR-1744011. SN-P acknowledges support from a KNI-Weathley fellowship. JA, GR, FvO, SN-P, and HR acknowledge support of IQIM (NSF funded physics frontiers center). JK acknowledges support from Deutsche Forschungsgemeinschaft (DFG 406557161), YC a Kwanjeong fellowship, FvO DFG support through CRC 183, and JA support from the NSF through grant DMR-1723367. YP, AT and JA are grateful to support from the Walter Burke Institute for Theoretical Physics at Caltech. **Author Contributions:** YC, JK, and SNP conceived the experiment. YC and JK preformed the measurements. YC made the samples with the help of HA, RP, and YZ. YC, JK, HR and SNP performed data analysis. YP and AT developed the theory guided by FvO, JA, and GR. YC, JK and SNP wrote the manuscript with input from all authors**. Competing interests**: Authors declare no competing financial interests. **Data and materials availability:** Data is available upon reasonable request.

**Figure Captions:**

**Figure 1: Twisted bilayer graphene (TBG). (A)** The twist angle $\theta$ in TBG gives rise to the formation of a triangular moiré lattice of regions corresponding to AA stacking (blue circles) with period $L_M$ separated by AB and BA stacking regions (see green circle and the corresponding inset). Large density of states is expected to be localized on the AA regions (*4–6*). **(B, C)** Schematics of the TBG band structure and the corresponding density of states showing vHs peaks. For angles close to $\theta$ **=** 1.1**°** correlations are expected to deform the vHs peaks (marked in red). **(D)** Schematic of the experiment: TBG is placed on top of an atomically smooth dielectric (boron nitride, BN) and back gated by an approximately 10nm thick graphite layer. **(E)** Optical image of the sample. Dashed areas correspond to two graphene layers. Scale bar corresponds to 20µm. **(F)** Topography showing the moiré pattern for θ=1.92˚ area of device D1. Note that $L_M$ is not identical for the different directions ($L_{MA}$ = 7.1nm; $L_{MB}$ = 7.6nm; $L_{MC}$ = 7.3nm) indicating small external strain in the sample. Set point conditions: $V_s$ = -200mV and $I_s$ = 50pA. **(G)** Topography for θ=1.01˚ area of device D2 ($L_{MA}$ = 13.8nm; $L_{MB}$ = 13.7nm; $L_{MC}$ = 14.2nm). Set point conditions: $V_s$ = 500mV and $I_s$ = 50pA. **(H, I)** Spectroscopy on an AA site for slight electron (H, $V_{Bg}$ = -0.6V) and hole (I, $V_{Bg}$ = -4.6V) doping induced by a back gate voltage $V_{Bg}$, for θ = 1.92˚. **(J, K)** Spectroscopy on an AA site for slight electron (J, $V_{Bg}$ = -0.7V) and hole (K, $V_{Bg}$ = -3.6V) doping close to the magic angle θ=1.01˚.

**Figure 2: Evolution of the TBG point spectrum with back-gate voltage. (A)** Point spectra on an AA site as a function of $V_{Bg}$, which tunes the overall charge density and therefore the position of the Fermi level. The data allow identification of different regimes indicated by the colored horizontal lines. Green – Fermi level in the dispersive bands; orange – Fermi level in the gap between the flat and dispersive band; red – Fermi level in the top flat band; gray – Fermi level in the bottom flat band. **(B)** Schematics of the TBG band structure with colors indicating the different regimes shown in **(A)**. **(C)** Gap between the electron flat band and the upper dispersive bands vs angle as extracted from similar data for different angles. **(D)** Line traces from **(A)** for fully occupied bands (red, $V_{Bg}$ = 9V) and close to the CNP (blue, $V_{Bg}$ = -3V). **(E)** Difference $\Delta_{CNP}$ between the combined bandwidth at the CNP and away from the CNP as a function of angle θ.

**Figure 3: Model calculations and breaking of $C_3$ symmetry. (A)** Density of states as a function of energy and filling for a single flavor ten band model with short range Coulomb interactions. The value of $E_c/W$=1 is chosen to qualitatively match the experiment upon a scaling factor of approximately 15. **(B, C)** Line cuts at the CNP and at filling $n_f$=-1 showing enhanced splitting at the CNP and broadening of the peaks compared to the non-interacting model. **(D)** Calculated band structure across the Brillouin zone showing the position of the Dirac points at the CNP. **(E)** Band structure obtained from the $C_3$-broken state for different filling factors (left to right $n_f$=-0.75, -0.5, -0.25, 0, 0.25, 0.5, 0.75). Upper (lower) panels correspond to the electron (hole) flat band. **(F)** Calculated local density of states (LDOS) profiles at the bridges connecting different AA sites, represented with lines of width directly proportional to the LDOS (see (*22*), section 5, for more details). **(G)** Spatially resolved TDOS maps near the CNP highlighting anisotropy at different bias voltages $V_{Bias}$ (with $V_{Bg}$ = -4.5V). From the upper left corner to the lower right: -18mV; -12mV, upper panels; -8mV; 0mV, central panels; 8mV; 12mV, lower panels.

**Figure 4: Spectroscopy at half-filling of the flat bands. (A)** Waterfall plot showing the evolution of point spectra as a function of $V_{Bg}$ around the CNP for an AA site in an area with θ=1.01˚. Line cuts corresponding to half filling on the electron and hole side are shown in **(B)** and **(C)**. **(D)** Conductance at the Fermi level ($V_{Bias}$=0mV) showing dips in the TDOS near half filling on both the electron and hole side as indicated by green and magenta rectangles. The black arrows indicate the expected positions corresponding to one quarter and three quarter filling. Blue areas correspond to the charge neutrality and the onset of the upper dispersive band.

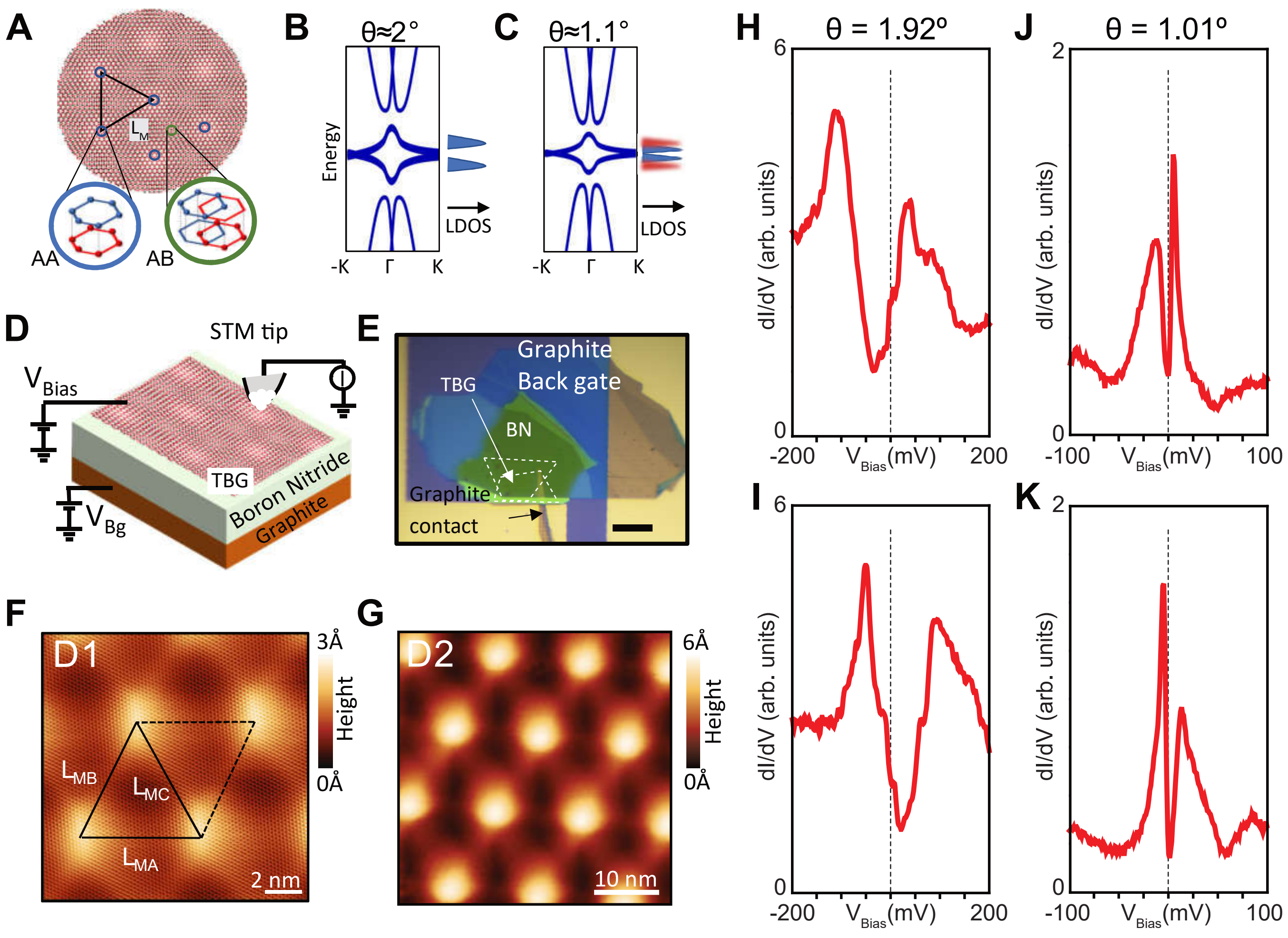

Figure 1

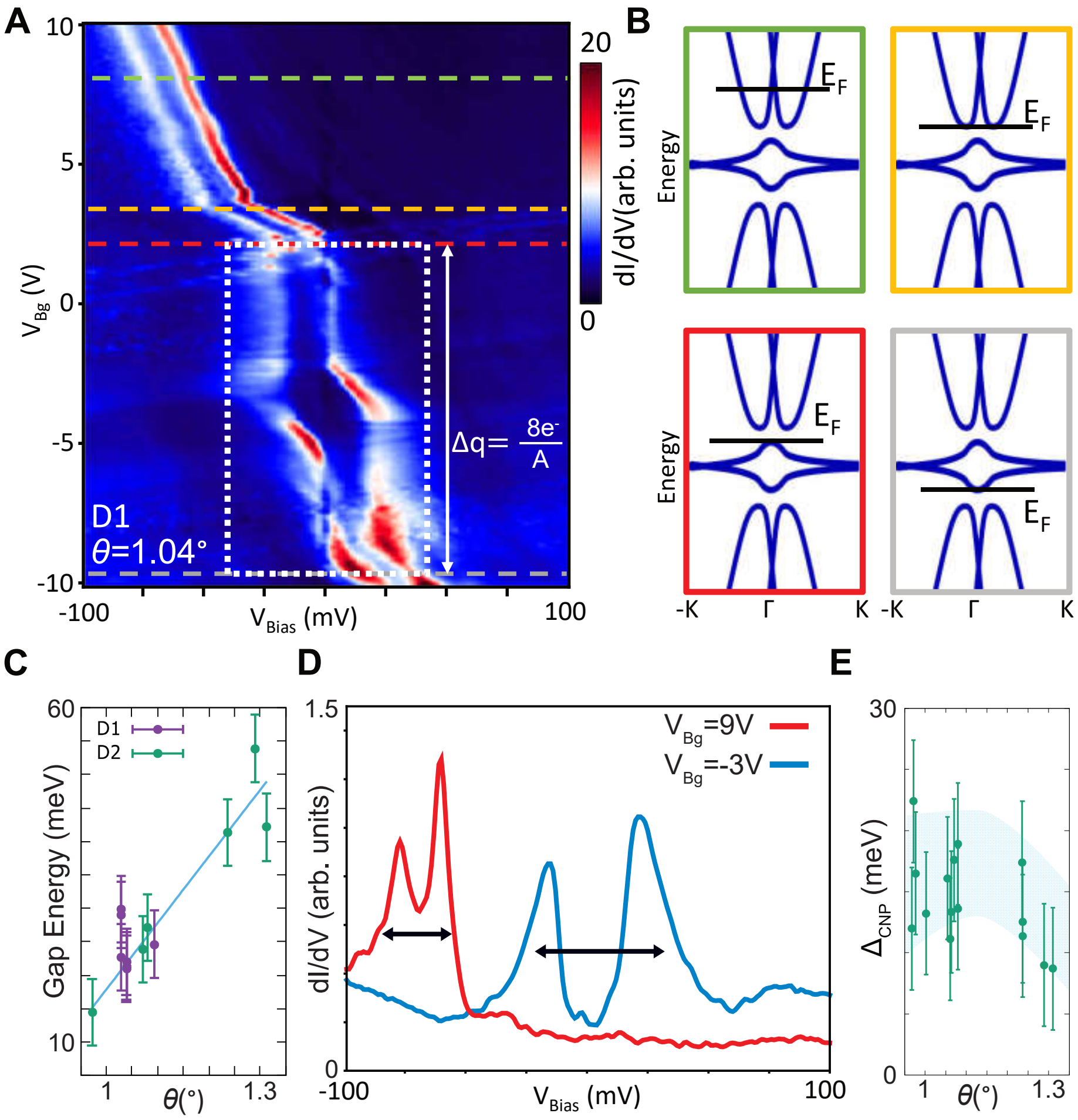

Figure 2

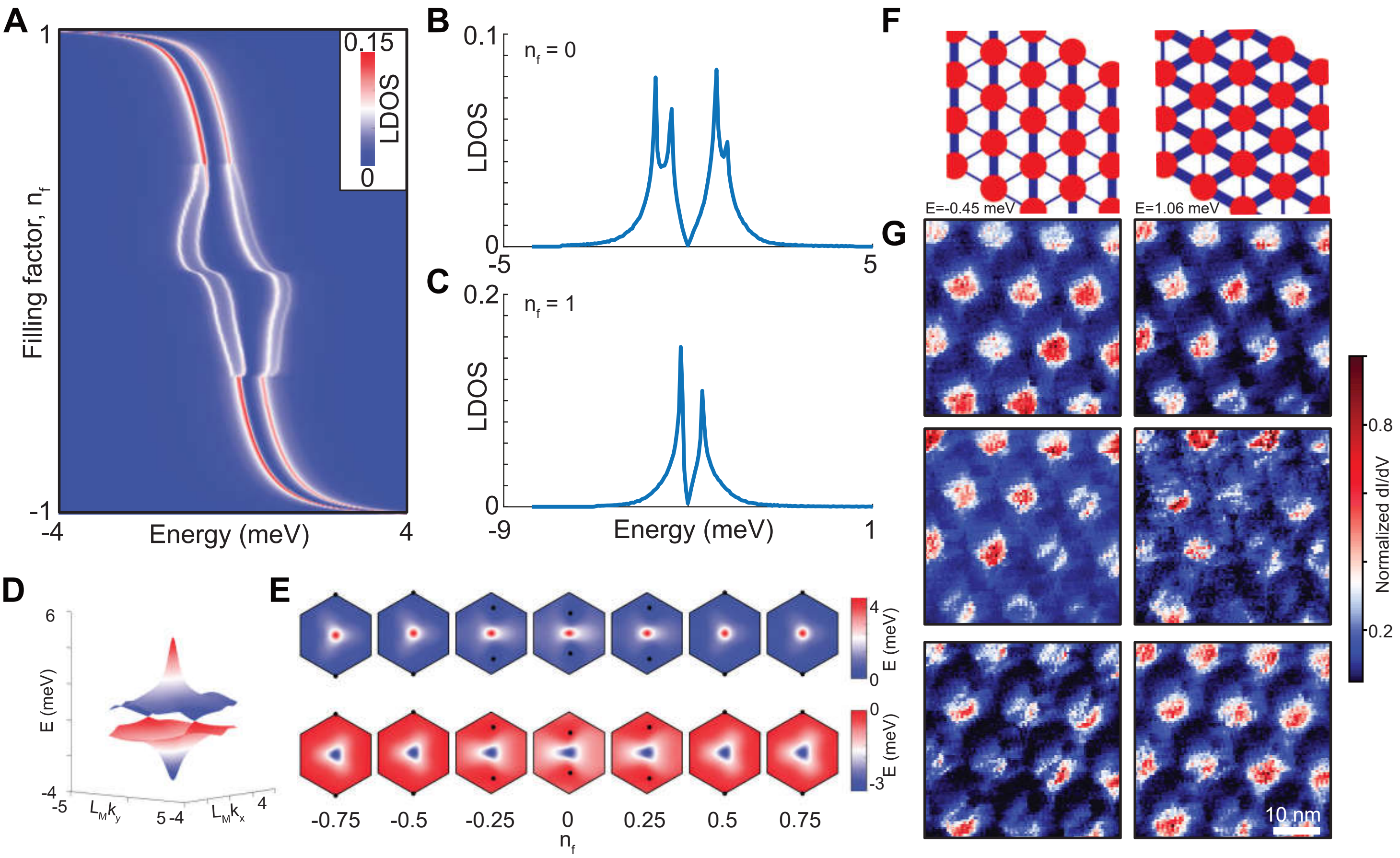

Figure 3

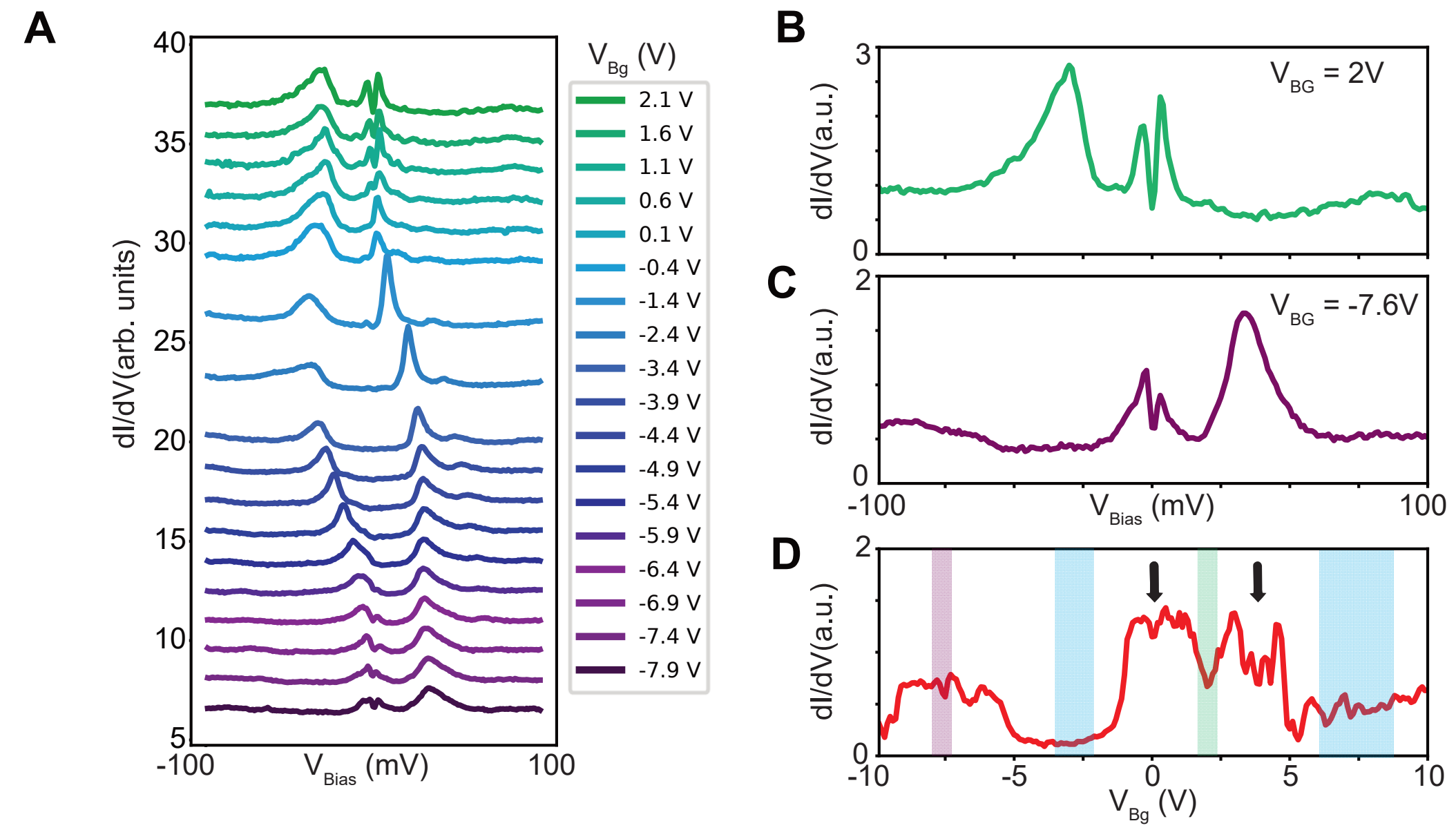

Figure 4

**Supplementary information for "Imaging Electronic Correlations in Twisted Bilayer Graphene Near the Magic Angle"**

**Authors:** Youngjoon Choi, Jeannette Kemmer, Yang Peng, Alex Thomson, Harpreet Arora, Robert Polski, Yiran Zhang, Hechen Ren, Jason Alicea, Gil Refael, Felix von Oppen, Kenji Watanabe, Takashi Taniguchi, and Stevan Nadj-Perge

Table of contents

**Materials and methods**

The measurements are performed in a Unisoku USM 1300J Scanning Tunneling Microscopy/Atomic Force Microscopy system in STM mode. We use a Pt/Ir tip prepared on a silver (Ag) crystal by observing quasiparticle interference and a spectrum that shows the Ag(111) surface state. Before approaching the TBG sample, we verify that the density of states of the tip is featureless in the range of +/-200mV by taking spectra of the gold electrodes before and after the measurements. We have measured in total five devices and three of them showed a moiré pattern with a period close to 13nm corresponding to a twist angle $\theta \approx 1.1^\circ$. Two of these devices were gate tunable in the range between +/- 10V. The effective electron temperature of the system T = 1.5K is calibrated on the superconducting gap of a lead Pb(110) crystal mounted in a similar measurement configuration as the twisted bilayer graphene. The lock-in parameters are: excitation voltage $V_{ac}$ = 400μV-1mV and frequency f = 433Hz. The scanner calibration and the topography data are verified on the Pb(110) crystal.

**Supplementary Text**

**Section 1 - Sample fabrication and measurement details**

Figure S1 outlines the fabrication steps for twisted bilayer graphene (TBG). We used the tear-and-twist technique following similar procedures outlined in Refs (*1*, *15*, *17*). After picking up a 30-50nm-thick boron nitride flake using Polydimethylsiloxane (PDMS) coated with a Poly(bisphenol A carbonate) (PC) polymer, the graphene is torn into two parts which were subsequently picked up while controlling the twist between the parts. The entire stack is then transferred onto a separate PDMS film in order to flip the order of the layers (Fig. S1 left panel). In this step, PC is dissolved in a N-Methyl-2-pyrrolidone (NMP). Afterwards the PDMS with the inverted stack structure is transferred onto a prepared silicon oxide chip with prepatterned electrodes and a 10nm thick metallic graphene multilayer that is used as the back gate (Fig. S1 right panel). The twisted bilayer graphene (TBG) is then contacted to the gold electrodes using additional few layer graphene contacts. During device fabrication special care is taken so that the temperature of the sample never exceeds 150˚C to avoid untwisting of the TBG (*1*).

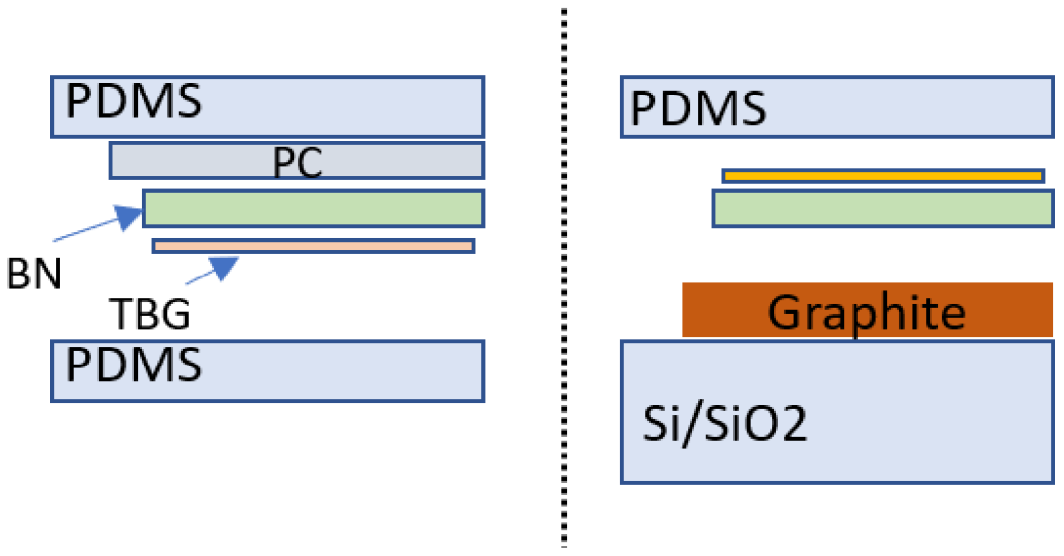

Figure S1. Critical steps of device fabrication sequence. After picking up the BN and graphene flakes, the stack is transferred and flipped onto a second PDMS stamp that is used for a second transfer onto the graphite back gate. In a separate step, a few-layer graphene is transferred to contact TBG with a bias electrode.

**Section 2 - Data on 4.8nm moiré pattern**

Fig. S2 shows a dI/dV map for the $L_M$= 4.8nm (device D2) moiré pattern. The charge neutrality point and both vHs peaks move in parallels following the quadratic relation between charge density $n$ and chemical potential $\mu$, $|n| = \alpha\, \mu^2$ (up to an offset in $V_{Bias}$ and $V_{Bg}$ ). This suggests that in this regime the density of states between the vHs peaks follows approximately a linear dispersion as expected for large twist angles. The vertical offset between the two branches is a consequence of tip-induced gating, which is explained in the following section. We note that this effect may explain the apparent small increase in the distance between the van Hove singularities at the CNP observed in Ref. (*14*). Importantly, the relative enhancement of the peak splitting for angles close to the magic angle value is too large to be explained by this effect.

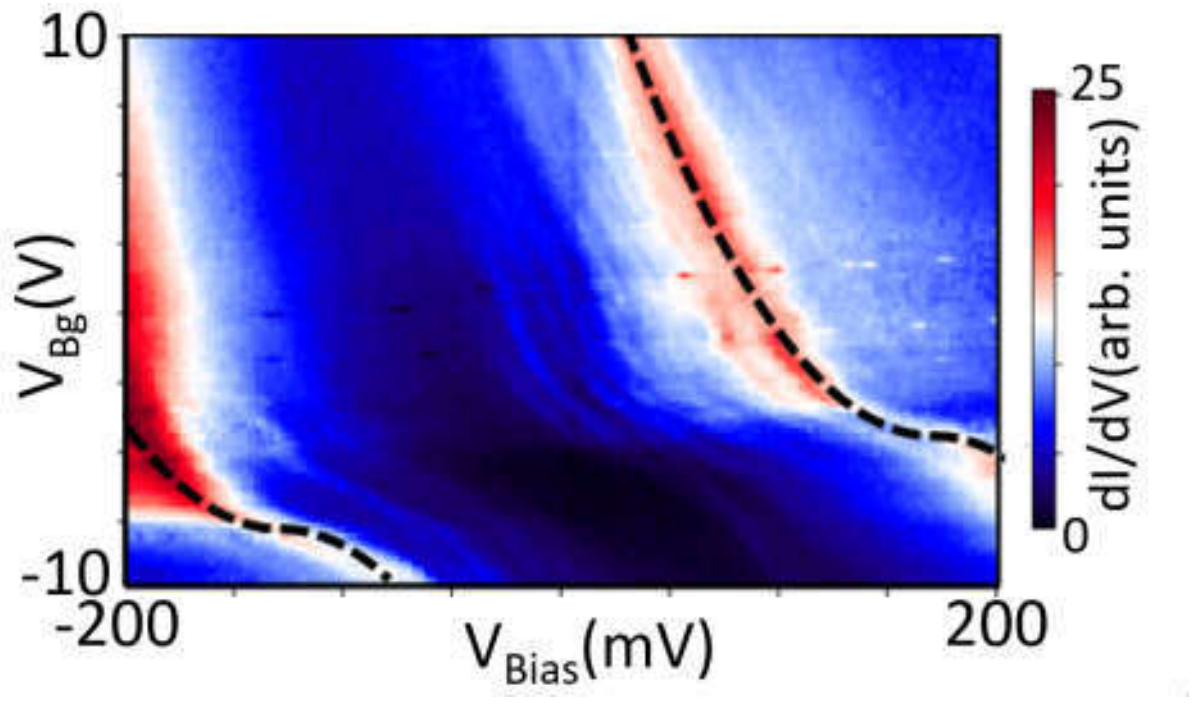

Figure S2. Spectroscopy of the AA sites for θ≈3°, device D2. Dashed lines correspond to a fit $|n|=\alpha\mu^2$ where μ is $\mu = -eV_{Bias}$ and $n$= $C_{Bg}$ $V_{Bg}$. The small vertical offset of the two branches that gives rise to a small enhancement in splitting in the $V_{Bias}$ direction is due to tip gating as discussed in following section.

## Section 3 - Tip related effects

### Dynamic strain effects

Occasionally we observe topographies with pronounced hysteresis, especially for parameters when the STM tip-sample distance is reduced (for example $V_{Bias}$=±200mV, set point current 100pA or more). Similar STM-tip-induced strain effects have been previously reported for moiré patterns in single layer graphene/ boron nitride heterostructures (*42*) as well as in graphene on silicon oxide (*43*). The intuitive understanding of this effect is that when the interaction between the tip and the sample is strong, the tip can slightly displace the carbon atoms as it scans over the surface and consequently shift the moiré pattern boundaries. We have performed a similar analysis as in Ref. (*42*) and have obtained strain maps illustrating this effect (see Fig. S3). In addition to these strain related effects, we have also observed other signatures of tip-graphene interaction, such as hysteresis in the current vs height relation I(z).

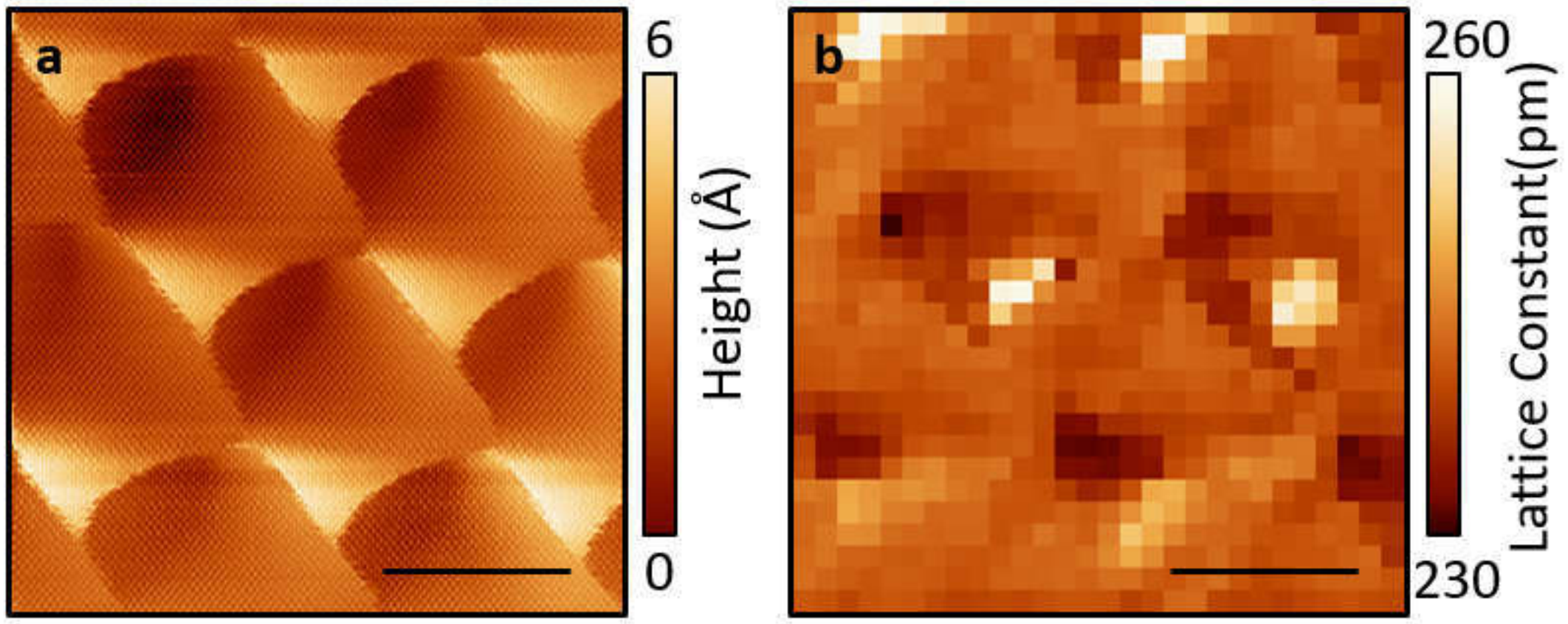

Figure S3. **(a)** Topography of the strained region close to the magic angle ($L_M$ = 13nm) in device D1 and the corresponding strain map **(b)**. The map represents the local lattice constant obtained by coarsening the topography grid into discrete pixels and performing a Fourier analysis of the 3nmx3nm area around each pixel and selecting one particular direction. Other directions show a similar dependence. Scale bar is 10nm for both plots.

The dynamic strain effects are not observed consistently throughout both samples. The strain is typically observed in areas close to the magic angle in device D1 and in certain areas away from the magic angle in device D2. Also in regions close to the magic angle in device D2, the moiré superlattice is more uniform and there is overall less strain. From these observations we conclude that the presence of intrinsic external strain is needed in order to observe the dynamic tip induced strain effect. Also different STM microtips may have different sensitivities to these effects. Importantly, we do not see a qualitative difference in the spectroscopy data between the two devices indicating that these effects do not play a significant role in determining the local density of states. This is consistent with the previously reported observations in Ref. (*42*). While this effect produces artifacts in the topography, it does not change it permanently so the distance between AA sites and hence the twist angle θ can still be obtained accurately.

**Effects of the tip screening**

Due to the close proximity between the STM tip and the sample, one might expect that the presence of metallic objects will facilitate screening of electronic correlations. In this section, we give estimates of the strength of the electron-electron interaction in this system. Indeed, in the presence of the metal tip which provides additional screening the interaction energy scale can be greatly suppressed compared to the pristine unscreened case. A naïve estimate for the electrostatic energy for two electrons placed $L_M$=13nm apart is given by $e^2/(4\pi\varepsilon\varepsilon_{BN}L_M)$ = 36meV. This estimate is enhanced further to approximately 50meV if we consider that we have the boron nitride dielectric only on one side. However, in the presence of a metallic tip this interaction is locally screened considerably. For a tip-sample distance of 1nm (taken as an upper limit), the Coulomb-energy scale would scale as $(1/L_M - 1/\sqrt{(L_M^2 + 4z_{tip}^2)})$ resulting in an interaction energy of 0.5meV for $L_M$=13nm. More precise estimates for the decay of the Coulomb interaction are presented in Figure S4 as obtained by electrostatic simulations that take into account the presence of the metallic tip and the surrounding dielectric. The values of ~ 10mV extracted from comparing our experimental results to the model calculations presented in the main text are consistent with these estimates.

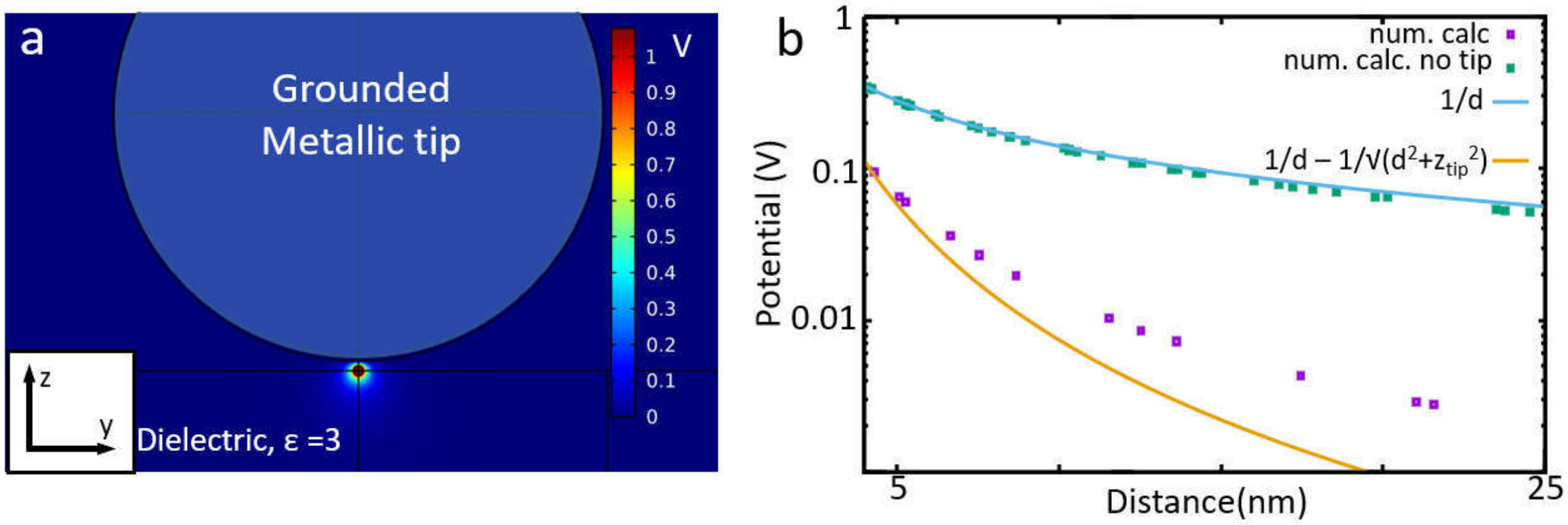

Figure S4. **(a)** 3D Electrostatic simulation of the potential that takes into account the tip sample geometry. Blue circle represents a metallic tip with the fixed potential V=0 and the rectangle corresponds to the dielectric slab. The small circle was charged by 1.6e-19C and the slice represents the decay of the potential along the y-z direction. **(b)** Decay of the potential as a function of distance. Different lines correspond to numerical calculations with and without the tip as well as theoretical estimates corresponding to 1/d decay and $(1/d-1/\sqrt{(d^2+4z_{tip}^2)})$ decay.

**Tip induced gating and work function difference between TBG and the tip**

It is established that the STM tip can change the local potential in semiconducting samples due to a finite screening length as observed in InAs (*44*) and monolayer and bilayer graphene in magnetic field (*40*) as well as other semiconducting systems. We have systematically observed the formation of quantum dots in the regions close to the magic angle indicative of the formation of insulating states. The induced quantum dots introduce a series of sharp resonances observed as almost horizontal lines crossing the features in $V_{Bias}$ vs $V_{Bg}$ conductance maps.

The observed resonance can be utilized to characterize the electrostatic properties of the quantum dot and determine the capacitance of the tip $C_{Tip}$ and the work function difference between the tip and the twisted bilayer graphene. In order to do this, we take measurements at different tip heights, i.e. different set points. Figure S5 shows a typical spectrum on an AA site for two different set currents of 100pA and 1nA. Lines (indicated by arrows) mark some of the resonances originating from the quantum dots. The slope of the lines directly measures the ratio between the tip and the back-gate capacitances $C_{Tip}/C_{Bg}$, and it is approximately 20 for device D2.

The tip capacitance changes when the tip moves closer to the sample. This is reflected in a change of the $V_{Bias}/V_{Bg}$ slope. Another effect observed is the overall shift of the positions in point spectroscopy $V_{Bias}$ vs. $V_{Bg}$ plots for which the upper flat band touches the Fermi level and also the position of charge neutrality. The shift of these points indicates that tip-induced-gating changes of the electron density underneath the tip, which is tip-sample distance dependent. This effect is a consequence of a difference between the work functions of the metallic tip and the twisted bilayer graphene.

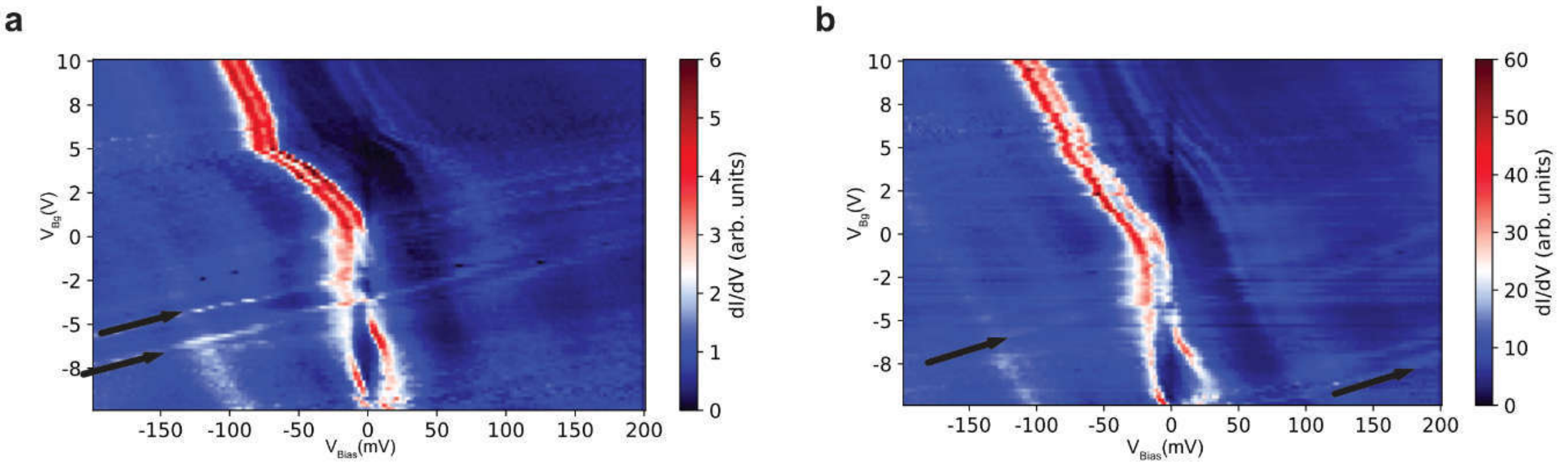

Figure S5: Spectroscopy of the AA site of device D2 for 100pA **(a)** and 1nA **(b)**. Black arrows indicate resonances originating due to tip-induced quantum dot**.** Note that the overall position of the point where the flat bands start crossing the Fermi level ($V_{Bias}$ = 0mV line) as well as the charge neutrality point shift towards more negative back gate voltages.

In a simple model, the charge density of the TBG underneath the tip can be written as:

$$n(r,z) = C_{Bg}\left(V_{Bg} - V_{Bias}\right) - C_{Tip}(r,z)(V_{Bias} - \Delta\Phi),$$

where $\Delta\Phi$ is the work function difference between the tip and the sample. Specifically, when the tip moves closer, the charge neutrality point moves towards more negative voltages ($V_{Bg}$=-6.8 V at CNP for 100pA setpoint and $V_{Bg}$=-8.6 V at CNP for 1nA). Also, the slope of the lines changes reflecting a change in the capacitance ($C_{Tip}/C_{Bg}$ = 20 for 100pA setpoint and 24 for 1nA setpoint). By solving the equation for the charge density at charge neutrality for two setpoints, one gets an estimate for $\Delta\Phi$ = 150-200mV. We note that the STM tips in our measurements are prepared on a silver crystal that has a smaller work function compared to graphene which results in the observed n-type doping.

**Section 4 - Areas with different twist angle θ**

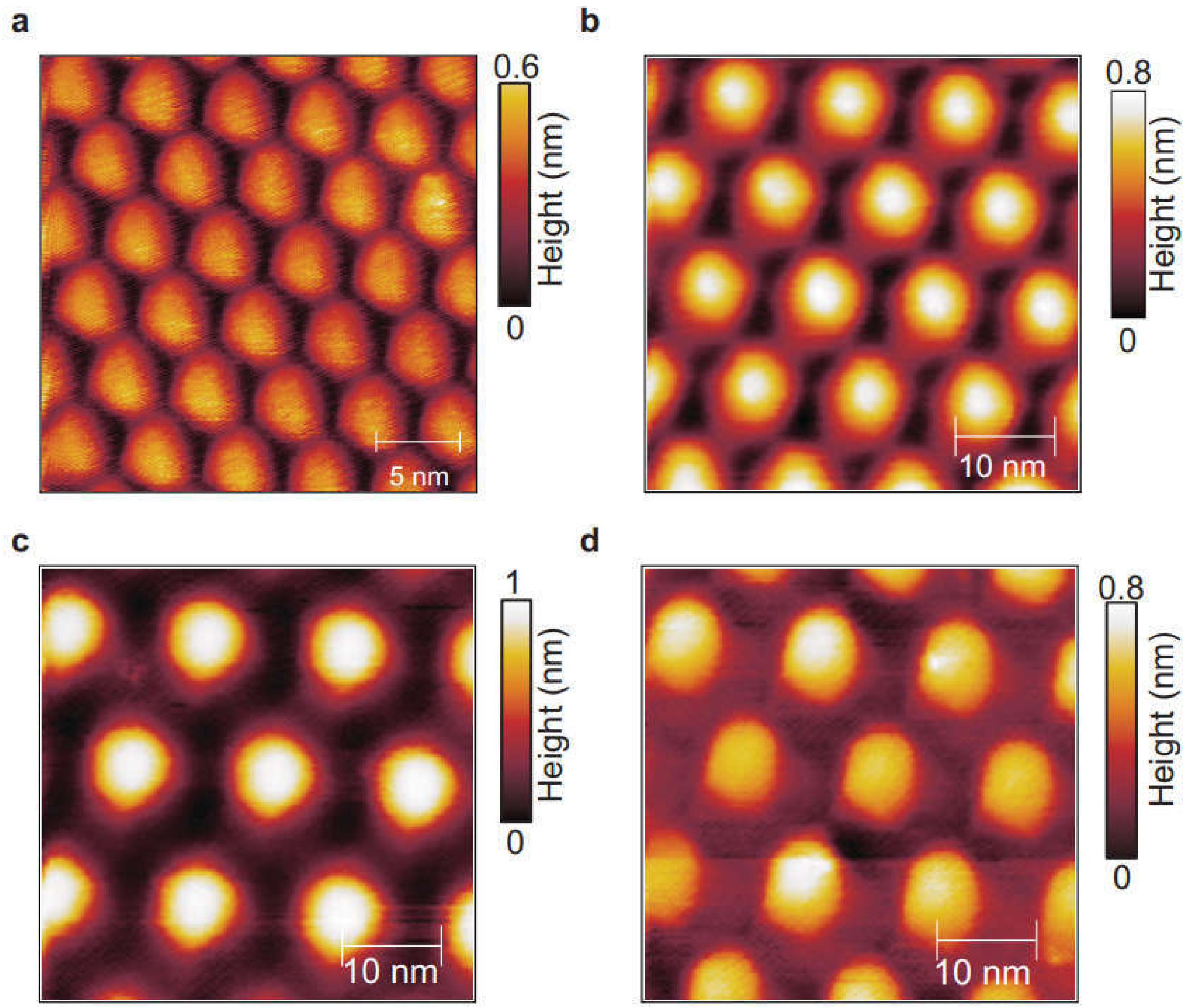

Figure S6. Examples of different areas of device D2 with the twist angle θ = 2.92° **(a)**; θ = 1.31° **(b)**; θ = 0.97° **(c)**; θ = 1.07° **(d)**. Set point conditions: $V_s$ = -200mV and $I_s$ = 30pA **(a,b)**; $I_s$ = 100pA **(c,d)**. In all samples we would typically find small clean areas with the lateral size of 50nm for which the angle θ is constant. Larger scale areas show significant amount of fabrication residues and strain.

## Section 5 - Theoretical Modeling

In this section we describe the theoretical model used to study the exchange effects in twisted bilayer graphene. We begin by reviewing the non-interacting Hamiltonian that serves as our starting point, before introducing Coulomb interactions and deriving a mean field Hamiltonian. We next present our results, and close with a brief discussion.

### A. Ten-band model for magic-angle bilayer graphene

Here, we briefly describe the ten-band model for magic-angle bilayer graphene, which was first introduced in Ref. (29). It has been shown (27,28) that the flat bands of magic-angle bilayer graphene cannot be captured by any minimal tight-binding model involving only the flat bands while retaining all symmetries. However, a tight-binding description is able to fully respect the symmetries of the system without fine-tuning when including additional bands.

In particular, the (effective) symmetries of the twisted bilayer graphene system are triangular lattice translations, $C_6$ rotations by $2\pi/6$, mirror symmetry $M_{2y}$ (which takes $y \to -y$, but actually represents a layer-exchanging 180° rotation in 3$d$ about the $x$-axis), and time-reversal $\mathcal{T}$.

Further, because the momentum difference between the $\boldsymbol{K}_{\text{lbz}}$ and $\boldsymbol{K}'_{\text{lbz}}$ points of the (large) Brillouin zone (denoted "lbz") of the microscopic graphene lattice is large compared to the moiré scale for small twist angles, states originating from each valley are effectively decoupled. This results in an effective U$_v$(1) valley symmetry, which enables us to focus on the electronic degrees of freedom in a single valley; the ten-band model of Ref. (29) takes advantage of this fact. However, given that both $C_6$ and $\mathcal{T}$ interchange the two valleys, only the combination $C_6\mathcal{T}$ is a proper symmetry of a single-valley theory. It follows that the spatial symmetries of the ten-band model are generated entirely by $C_2\mathcal{T}$, $C_3$, and $M_{2y}$.

Finally, the flat bands we wish to describe possess two Dirac cones (per spin, per valley) at the $\boldsymbol{K}$ and $\boldsymbol{K}'$ points of the (moiré) Brillouin zone. (We reserve $\boldsymbol{K}$, $\boldsymbol{K}'$, $\boldsymbol{\Gamma}$, etc. to describe the points of the small moiré BZ.) The masslessness of the cones is ensured by $C_2\mathcal{T}$ symmetry, while their location at $\boldsymbol{K}$ and $\boldsymbol{K}'$ is protected by $C_3$.

Before introducing interactions between electrons, we note that an overall scaling factor between the ten-band model and the actual physical system must be included. This factor can be obtained by comparing the flat bandwidth of the ten-band model and the separation of van Hove singularities measured in the tunneling DOS in the energy regime far away from the CNP, where electron-interaction effects are presumably less important. We estimate this factor to be around $15$. In all that follows, we keep the units set by the ten-band model, only scaling by $\sim 15$ when comparing against experiment.

### 1. *Non-interacting Hamiltonian*

The ten-band model is defined on a triangular lattice with basis vectors $\boldsymbol{a}_1 = (\sqrt{3}/2, -1/2)$ and $\boldsymbol{a}_2 = (0,1)$. We write the Bravais lattice sites as $\boldsymbol{r} = r_1\boldsymbol{a}_1 + r_2\boldsymbol{a}_2$ or simply as $\boldsymbol{r} = (r_1, r_2)$, where $r_{1,2} \in \mathbb{Z}$. Within each unit cell, there are ten orbitals which are distributed on three different sites, as indicated by the different colors in Fig. S7. Explicitly, there are three orbitals, $p_z$, $p_+$, and $p_-$, on every triangular lattice site (red). Each of the three kagome sites (black) within a unit cell hosts an $s$ orbital. Finally, both A and B sublattices of the honeycomb sites (blue) have $p_+$ and $p_-$ orbitals. Throughout this work, these ten orbitals are ordered as $c_{\boldsymbol{r}} = (\tau_{z,\boldsymbol{r}}, \tau_{+,\boldsymbol{r}}, \tau_{-,\boldsymbol{r}}, \kappa_{1,\boldsymbol{r}}, \kappa_{2,\boldsymbol{r}}, \kappa_{2,\boldsymbol{r}}, \eta_{A+,\boldsymbol{r}}, \eta_{A-,\boldsymbol{r}}, \eta_{B+,\boldsymbol{r}}, \eta_{B-,\boldsymbol{r}})^T$, where $\tau$, $\kappa$, and $\eta$ respectively denote operators on the triangular, kagome, and honeycomb sites.

The Bloch Hamiltonian of the ten-band model can be written as

$$\mathcal{H} = \sum_{\boldsymbol{k}} \sum_{i,j=1}^{10} c^\dagger_{i\boldsymbol{k}} H_{ij}(\boldsymbol{k}) c_{j\boldsymbol{k}}, \tag{S1}$$

where $\boldsymbol{k}$ is the Bloch momentum, $i, j$ are the orbital indices, and $c_j(\boldsymbol{k})$ annihilates an electron on orbital $j$ at momentum $\boldsymbol{k}$. In our conventions the real-space and momentum-space electron operators are related via $c_{i,\boldsymbol{r}} = \sum_{\boldsymbol{k}} \exp(i\boldsymbol{k}\cdot\boldsymbol{r}) c_i(\boldsymbol{k})/\sqrt{\mathcal{V}}$, with $\mathcal{V}$ the area of the system.

The 10×10 Hamiltonian matrix $H(\boldsymbol{k})$ is defined as a sum of two terms

$$H(\boldsymbol{k}) = H_0(\boldsymbol{k}) + V(\boldsymbol{k}). \tag{S2}$$

The first term, $H_0(\boldsymbol{k})$, describes hoppings from the honeycomb sites to the triangular and kagome sites; this piece takes the form

$$H_0(\boldsymbol{k}) = t \begin{pmatrix} 0_{6\times 6} & h(\boldsymbol{k}) \\ h(\boldsymbol{k})^\dagger & 0_{4\times 4} \end{pmatrix}, \tag{S3}$$

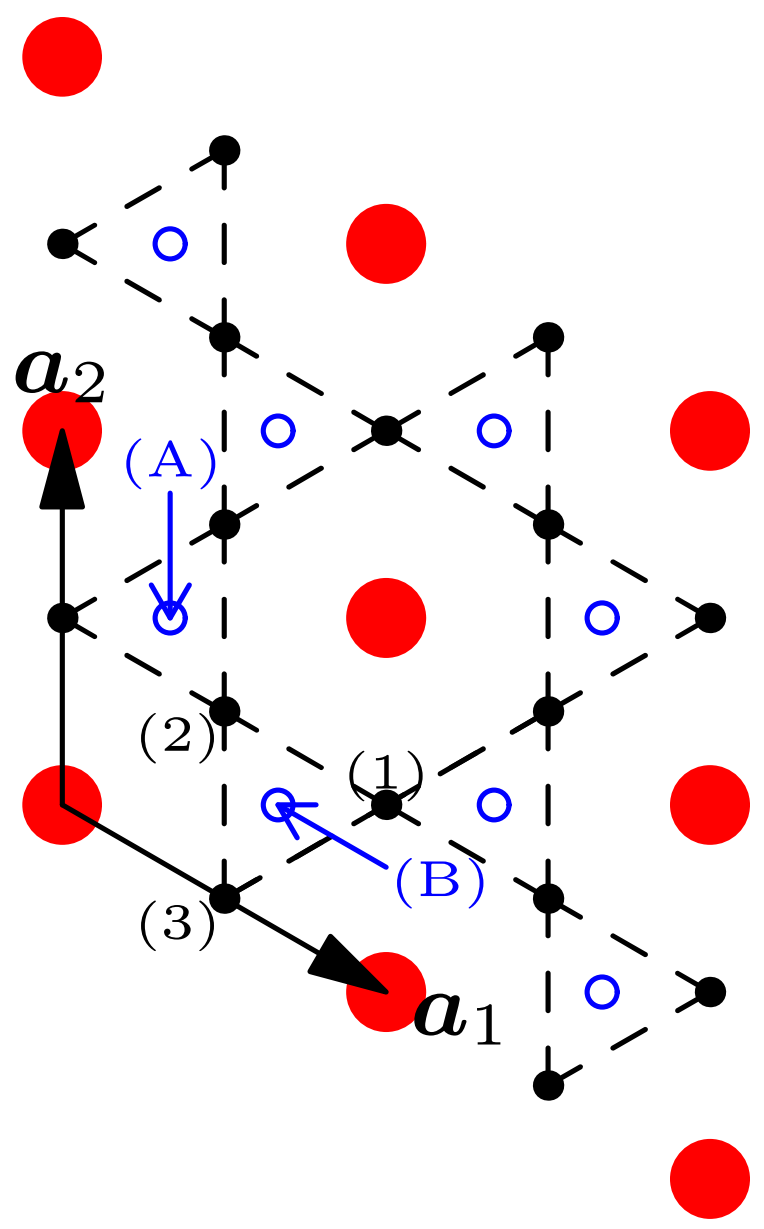

Figure S7. Lattice and orbitals for the ten-band model. The red solid circles denote the triangular sites with $p_z$, $p_+$, and $p_-$ orbitals. The black solid circles correspond to the three types of kagome sites, labeled as (1), (2), and (3). On each of these kagome sites, there is an $s$ orbital. The blue empty circles indicate the honeycomb sites, either type A or type B, with $p_+$ and $p_-$ orbitals on each of them.

with

$$h(\boldsymbol{k}) = \begin{pmatrix} (1+\phi_{11}\omega^*+\phi_{01}\omega)\zeta^* a & -(\omega+\phi_{11}\omega^*+\phi_{01})\zeta^* a & -(\omega+\phi_{10}+\phi_{11}\omega^*)\zeta a & (\omega+\phi_{10}\omega^*+\phi_{11})\zeta a \\ (1+\phi_{11}+\phi_{01})c & (\omega^*+\phi_{11}\omega+\phi_{01})\zeta b & (1+\phi_{10}+\phi_{11})c & (\omega^*+\phi_{10}\omega+\phi_{11})\zeta^* b \\ (1+\omega\phi_{11}+\omega^*\phi_{01})\zeta b & (1+\phi_{11}+\phi_{01})c & (\omega^*+\phi_{10}+\phi_{11}\omega)\zeta^* b & (1+\phi_{10}+\phi_{11})c \\ -i\phi_{\bar{1}0}d & -i\phi_{\bar{1}0}d & id & id \\ -i\omega^* d & -i\omega d & i\omega^* d & i\omega d \\ -i\phi_{01}\omega d & -i\phi_{01}\omega^* d & i\omega d & i\omega^* d \end{pmatrix}, \tag{S4}$$

where $\phi_{\ell m} = \exp(-i(\ell k_1 + m k_2)$, $k_i = \boldsymbol{k}\cdot\boldsymbol{a}_i$, $\bar{\ell} = -\ell$, and $\omega = \exp(i2\pi/3)$.

The remaining tunneling amplitudes are included in the second term, $V(\boldsymbol{k})$, which can be written as

$$V(\boldsymbol{k}) = \begin{pmatrix} \mu_{p_z} & \mathcal{C}^\dagger_{p_\pm p_z} & 0 & 0 \\ \mathcal{C}_{p_\pm p_z} & H_{p_\pm} + \mu_{p_\pm}\mathbb{I}_{2\times 2} & \mathcal{C}^\dagger_{\kappa p_z} & 0 \\ 0 & \mathcal{C}_{\kappa p_z} & \mu_\kappa \mathbb{I}_{3\times 3} & 0 \\ 0 & 0 & 0 & H_\eta \end{pmatrix}. \tag{S5}$$

Table I. Parameters for the ten-band model.

| $t_0$ | $a$ | $b$ | $c$ | $d$ | $t_\eta$ | $\mu_{p_z}$ | $t_{p_\pm}$ | $t_{p_\pm p_\pm}$ | $\mu_{p_\pm}$ | $\mu_\kappa$ | $t_{p_\pm p_z}$ | $t^+_{\kappa p_\pm}$ | $t^-_{\kappa p_\pm}$ |
|---|---|---|---|---|---|---|---|---|---|---|---|---|---|
| 130.0 meV | 0.110 | 0.033 | 0.033 | 0.573 | 32.5 meV | $-0.1t_\eta$ | $0.003t_\eta$ | $0.004t_\eta$ | $3t_{p_\pm}$ | $0.11t_\eta$ | $0.016t_\eta$ | $0.016t_\eta$ | $-0.016t_\eta$ |

Here $\mathbb{I}$ denotes the identity matrix, and we have defined

$$H_{p_\pm} = \begin{pmatrix} t_{p_\pm}(\phi_{01}+\phi_{11}+\phi_{10}+h.c.) & t_{p_\pm p_\pm}(\phi_{01}+\omega^*\phi_{\bar{1}\bar{1}}+\omega\phi_{10}) \\ t_{p_\pm p_\pm}(\phi_{0\bar{1}}+\omega\phi_{11}+\omega^*\phi_{\bar{1}0}) & t_{p_\pm}(\phi_{01}+\phi_{11}+\phi_{10}+h.c.) \end{pmatrix} \tag{S6}$$

to describe the $p_\pm$ orbitals on the triangular sites and

$$H_\eta = t_\eta \begin{pmatrix} 0 & i(1+\phi_{0\bar{1}}+\phi_{10}) \\ -i(1+\phi_{01}+\phi_{\bar{1}0}) & 0 \end{pmatrix} \otimes \mathbb{I}_{2\times 2} \tag{S7}$$

to describe the honeycomb orbitals. Moreover, the coupling between the $p_\pm$ and $p_z$ orbitals on the triangular sites, and the coupling between $s$ orbitals of the kagome sites and $p_\pm$ of the triangular sites, are respectively described by

$$\mathcal{C}_{p_\pm p_z} = it_{p_\pm p_z} \begin{pmatrix} \phi_{01}+\phi_{\bar{1}\bar{1}}\omega+\phi_{10}\omega^* \\ -(\phi_{0\bar{1}}+\phi_{11}\omega^*+\phi_{\bar{1}0}\omega) \end{pmatrix}, \tag{S8}$$

and

$$\mathcal{C}_{\kappa p_\pm} = t^+_{\kappa p_\pm} \begin{pmatrix} \phi_{\bar{1}0} & \phi_{\bar{1}\bar{1}} \\ \phi_{\bar{1}\bar{1}}\omega^* & \omega \\ \omega & \phi_{\bar{1}0}\omega^* \end{pmatrix} - t^-_{\kappa p_\pm} \begin{pmatrix} \phi_{\bar{1}\bar{1}} & \phi_{\bar{1}0} \\ \omega^* & \phi_{\bar{1}\bar{1}}\omega \\ \phi_{\bar{1}0}\omega & \omega^* \end{pmatrix}. \tag{S9}$$

Table I lists the parameters for the ten-band model.

### 2. *Symmetries*

As we discussed above, the model introduced in the previous section preserves the symmetries twisted bilayer graphene should possess at small angles: $C_2\mathcal{T}$, $C_3$, and $M_{2y}$, In the following, we explicitly describe the action of these on the degrees of freedom in our model.

*a.* $C_2\mathcal{T}$*:* Under $C_2\mathcal{T}$, the basis vectors transform as

$$C_2\mathcal{T}: \quad \boldsymbol{a}_1 \to -\boldsymbol{a}_1, \quad \boldsymbol{a}_2 \to -\boldsymbol{a}_2, \quad i \to -i. \tag{S10}$$

It follows that the orbitals become

$$
C_2\mathcal{T}: \qquad \begin{pmatrix} \tau_{z,\boldsymbol{r}} \\ \tau_{+,\boldsymbol{r}} \\ \tau_{-,\boldsymbol{r}} \end{pmatrix} \to \begin{pmatrix} \tau_{z,\boldsymbol{r}'} \\ \tau_{-,\boldsymbol{r}'} \\ \tau_{+,\boldsymbol{r}'} \end{pmatrix}, \qquad \begin{pmatrix} \kappa_{1,\boldsymbol{r}} \\ \kappa_{2,\boldsymbol{r}} \\ \kappa_{3,\boldsymbol{r}} \end{pmatrix} \to \begin{pmatrix} \kappa_{1,\boldsymbol{r}'-2\boldsymbol{a}_1-\boldsymbol{a}_2} \\ \kappa_{2,\boldsymbol{r}'-\boldsymbol{a}_1-\boldsymbol{a}_2} \\ \kappa_{3,\boldsymbol{r}'-\boldsymbol{a}_1} \end{pmatrix},
$$

$$
\begin{pmatrix} \eta_{A+,\boldsymbol{r}} \\ \eta_{A-,\boldsymbol{r}} \end{pmatrix} \to \begin{pmatrix} \eta_{B-,\boldsymbol{r}'-\boldsymbol{a}_1-\boldsymbol{a}_2} \\ \eta_{B+,\boldsymbol{r}'-\boldsymbol{a}_1-\boldsymbol{a}_2} \end{pmatrix}, \qquad \begin{pmatrix} \eta_{B+,\boldsymbol{r}} \\ \eta_{B-,\boldsymbol{r}} \end{pmatrix} \to \begin{pmatrix} \eta_{A-,\boldsymbol{r}'-\boldsymbol{a}_1-\boldsymbol{a}_2} \\ \eta_{A+,\boldsymbol{r}'-\boldsymbol{a}_1-\boldsymbol{a}_2} \end{pmatrix}. \tag{S11}
$$

with $\boldsymbol{r}' = (-r_1, -r_2)$. In momentum space, we find

$$
C_2\mathcal{T}: \qquad c_{\boldsymbol{k}} \to M_{C_2\mathcal{T}}(\boldsymbol{k}) c_{\boldsymbol{k}}, \qquad\qquad i \to -i \tag{S12}
$$

where the unitary matrix $M_{C_2\mathcal{T}}(\boldsymbol{k})$ is given by

$$
M_{C_2\mathcal{T}}(\boldsymbol{k}) = \begin{pmatrix}
1 & & & & & & & & & \\
& 0 & 1 & & & & & & & \\
& 1 & 0 & & & & & & & \\
& & & \phi_{21} & & & & & & \\
& & & & \phi_{11} & & & & & \\
& & & & & \phi_{10} & & & & \\
& & & & & & 0 & 0 & 0 & \phi_{11} \\
& & & & & & 0 & 0 & \phi_{11} & 0 \\
& & & & & & 0 & \phi_{11} & 0 & 0 \\
& & & & & & \phi_{11} & 0 & 0 & 0
\end{pmatrix}. \tag{S13}
$$

*b.* *$C_3$:* Under rotation by $2\pi/3$, the basis vectors transform as

$$
C_3: \qquad \boldsymbol{a}_1 \to \boldsymbol{a}_2, \qquad\qquad \boldsymbol{a}_2 \to -\boldsymbol{a}_1 - \boldsymbol{a}_2. \tag{S14}
$$

We can show that $C_3$ takes the electron annihilation operators to

$$
C_3: \qquad \begin{pmatrix} \tau_{z,\boldsymbol{r}} \\ \tau_{+,\boldsymbol{r}} \\ \tau_{-,\boldsymbol{r}} \end{pmatrix} \to \begin{pmatrix} \tau_{z,\boldsymbol{r}'} \\ \omega^* \tau_{+,\boldsymbol{r}'} \\ \omega \tau_{-,\boldsymbol{r}'} \end{pmatrix}, \qquad \begin{pmatrix} \kappa_{1,\boldsymbol{r}} \\ \kappa_{2,\boldsymbol{r}} \\ \kappa_{3,\boldsymbol{r}} \end{pmatrix} \to \begin{pmatrix} \kappa_{2,\boldsymbol{r}'-\boldsymbol{a}_1} \\ \kappa_{3,\boldsymbol{r}'-\boldsymbol{a}_1} \\ \kappa_{1,\boldsymbol{r}'-\boldsymbol{a}_1} \end{pmatrix},
$$

$$
\begin{pmatrix} h_{A+,\boldsymbol{r}} \\ h_{A-,\boldsymbol{r}} \end{pmatrix} \to \begin{pmatrix} \omega^* h_{A+,\boldsymbol{r}'-\boldsymbol{a}_1-\boldsymbol{a}_2} \\ \omega h_{A-,\boldsymbol{r}'-\boldsymbol{a}_1-\boldsymbol{a}_2} \end{pmatrix}, \qquad \begin{pmatrix} h_{B+,\boldsymbol{r}} \\ h_{B-,\boldsymbol{r}} \end{pmatrix} \to \begin{pmatrix} \omega^* h_{B+,\boldsymbol{r}'-\boldsymbol{a}_1} \\ \omega h_{B-,\boldsymbol{r}'-\boldsymbol{a}_1} \end{pmatrix} \tag{S15}
$$

with $\boldsymbol{r}' = (-r_2, r_1 - r_2)$. This can be summarized in momentum space as

$$C_3 : \qquad c_{\boldsymbol{k}} \to M_{c_3}(\boldsymbol{k}) c_{\boldsymbol{k}'}, \qquad\qquad \boldsymbol{k}' = (-k_1 - k_2, k_1). \tag{S16}$$

where $\boldsymbol{k} = (k_1, k_2)$, $k_i = \boldsymbol{k} \cdot \boldsymbol{a}_i$ and

$$M_{C_3}(\boldsymbol{k}) = \begin{pmatrix} 1 &&&&&&&&& \\ & \omega^* &&&&&&&& \\ && \omega &&&&&&& \\ &&& 0 & \phi_{11}^* & 0 &&&& \\ &&& 0 & 0 & \phi_{11}^* &&&& \\ &&& \phi_{11}^* & 0 & 0 &&&& \\ &&&&&& \omega^*\phi_{01}^* &&& \\ &&&&&&& \omega\phi_{01}^* && \\ &&&&&&&& \omega^*\phi_{11}^* & \\ &&&&&&&&& \omega\phi_{11}^* \end{pmatrix}, \tag{S17}$$

with $\phi_{\ell m} = e^{-i(\ell k_1 + m k_2)}$.

*c.* $M_{2y}$*:* The mirror symmetry transforms of the lattice vectors are

$$M_{2y} : \qquad \boldsymbol{a}_1 \to \boldsymbol{a}_1 + \boldsymbol{a}_2, \qquad\qquad \boldsymbol{a}_2 \to -\boldsymbol{a}_2. \tag{S18}$$

We can show that this acts on the orbitals of the 10-band model as

$$M_{2y} : \quad \begin{pmatrix} \tau_{z,\boldsymbol{r}} \\ \tau_{+,\boldsymbol{r}} \\ \tau_{-,\boldsymbol{r}} \end{pmatrix} \to \begin{pmatrix} -\tau_{z,\boldsymbol{r}'} \\ \tau_{-,\boldsymbol{r}'} \\ \tau_{+,\boldsymbol{r}'} \end{pmatrix}, \qquad \begin{pmatrix} \kappa_{1,\boldsymbol{r}} \\ \kappa_{2,\boldsymbol{r}} \\ \kappa_{3,\boldsymbol{r}} \end{pmatrix} \to \begin{pmatrix} \kappa_{1,\boldsymbol{r}'} \\ \kappa_{3,\boldsymbol{r}'} \\ \kappa_{2,\boldsymbol{r}'} \end{pmatrix},$$
$$\begin{pmatrix} h_{A+,\boldsymbol{r}} \\ h_{A-,\boldsymbol{r}} \end{pmatrix} \to \begin{pmatrix} h_{A-,\boldsymbol{r}'-\boldsymbol{a}_2} \\ h_{A+,\boldsymbol{r}'-\boldsymbol{a}_2} \end{pmatrix}, \qquad \begin{pmatrix} h_{B+,\boldsymbol{r}} \\ h_{B-,\boldsymbol{r}} \end{pmatrix} \to \begin{pmatrix} h_{B-,\boldsymbol{r}'} \\ h_{B+,\boldsymbol{r}'} \end{pmatrix} \tag{S19}$$

with $\boldsymbol{r}' = (r_1, r_1 - r_2)$. In momentum space, this looks like

$$M_{2y} : \qquad c_{\boldsymbol{k}} \to M_{M_{2y}}(\boldsymbol{k}) c_{\boldsymbol{k}'}, \qquad\qquad \boldsymbol{k}' = (k_1 + k_2, -k_2), \tag{S20}$$

where

$$M_{M_{2y}}(\boldsymbol{k}) = \begin{pmatrix} -1 & & & & & & & & & \\ & 0 & 1 & & & & & & & \\ & 1 & 0 & & & & & & & \\ & & & 1 & & & & & & \\ & & & & 0 & 1 & & & & \\ & & & & 1 & 0 & & & & \\ & & & & & & 0 & \phi_{01}^* & & \\ & & & & & & \phi_{01}^* & 0 & & \\ & & & & & & & & 0 & 1 \\ & & & & & & & & 1 & 0 \end{pmatrix}. \quad \text{(S21)}$$

### B. Local Coulomb interaction and mean-field theory

We now supplement the ten-band model with symmetry-preserving electron-electron interactions, the form of which is determined by several factors. Motivated by the presence of screening from the STM tip, we consider a minimal short-range interaction. Microscopic calculations demonstrate that the flat-band wavefunctions are primarily weighted on the triangular-lattice "AA" sites of the moiré pattern (4,7,17). Accordingly, Po *et al.* (29) demonstrate that the majority of the spectral weight lies on the triangular lattice sites in the ten-band model. In spite of this, the overall charge density computed from more microscopically faithful models is nearly uniform (25). We therefore consider an on-site Coulomb interaction of the form $(n_{\boldsymbol{r}} - \langle n_{\boldsymbol{r}} \rangle)^2$, where $n_{\boldsymbol{r}}$ is the electron density and $\langle n_{\boldsymbol{r}} \rangle$ is the uniform background. Finally, since we are interested in correlation physics arising from partial filling of the flat bands, for simplicity we restrict the on-site Coulomb interaction to the $\tau$-orbitals:

$$H_C = E_C \sum_{\boldsymbol{r}} \left( \sum_{\mu = z, \pm} \Big[ n_{\mu, \boldsymbol{r}} - \langle n_{\mu, \boldsymbol{r}} \rangle \Big] \right)^2 \quad \text{(S22)}$$

where $n_{\mu,\boldsymbol{r}} = \tau_{\mu,\boldsymbol{r}}^\dagger \tau_{\mu,\boldsymbol{r}}$. We will consider neither valley nor spin symmetry breaking, so additional flavour indices need not be included.

The magnitude of the interaction $E_C$ is the only free parameter in our theory aside from the overall scaling factor needed to match the ten band model with experimental data in the weakly-interacting regime. We set $E_C = 1$ meV since it returns results reasonably similar to the experimental DOS shown in the main text.

To proceed, we perform a Hartree-Fock mean-field decoupling of the interaction $H_C$. By virtue of subtracting the uniform density in our definition of $H_C$, the Hartree term is completely cancelled, leaving only a Fock Hamiltonian:

$$\begin{aligned} H_{C,MF} &= -2E_C \sum_{\boldsymbol{r}} \sum_{\mu,\nu=z,\pm} \left\langle \tau^\dagger_{\nu,\boldsymbol{r}} \tau_{\mu,\boldsymbol{r}} \right\rangle \tau^\dagger_{\mu,\boldsymbol{r}} \tau_{\nu,\boldsymbol{r}} + E_C \sum_{\boldsymbol{r}} \sum_{\mu,\nu=z,\pm} \left\langle \tau^\dagger_{\nu,\boldsymbol{r}} \tau_{\mu,\boldsymbol{r}} \right\rangle \left\langle \tau^\dagger_{\mu,\boldsymbol{r}} \tau_{\nu,\boldsymbol{r}} \right\rangle \\ &= -2E_C \sum_{\boldsymbol{r}} \sum_{i,j=1}^{10} L_{ij} \left\langle c^\dagger_{i\boldsymbol{r}} c_{j\boldsymbol{r}} \right\rangle c^\dagger_{j\boldsymbol{r}} c_{i\boldsymbol{r}} + E_C \sum_{\boldsymbol{r}} \sum_{i,j=1}^{10} L_{ij} \left\langle c^\dagger_{i\boldsymbol{r}} c_{j\boldsymbol{r}} \right\rangle \left\langle c^\dagger_{j\boldsymbol{r}} c_{i\boldsymbol{r}} \right\rangle, \end{aligned} \tag{S23}$$

where we have defined the $10 \times 10$ matrix

$$L = \begin{pmatrix} 1 & 1 & 1 & \\ 1 & 1 & 1 & \\ 1 & 1 & 1 & \\ & & & 0_{7\times 7} \end{pmatrix}. \tag{S24}$$

It is convenient to write this in momentum space as

$$\begin{aligned} H_{C,MF} &= \sum_{ij=1}^{10} c^\dagger_{i\boldsymbol{k}} W_{ij} c_{j\boldsymbol{k}} - \frac{1}{2\mathcal{V}} \sum_{\boldsymbol{k}} \left\langle c^\dagger_{i\boldsymbol{k}} W_{ij}(\boldsymbol{k}) c_{j\boldsymbol{k}} \right\rangle \\ W_{ij} &= -\frac{2E_c}{\mathcal{V}} \sum_{\boldsymbol{k}} \langle c^\dagger_{j\boldsymbol{k}} c_{i\boldsymbol{k}} \rangle L_{ij}. \end{aligned} \tag{S25}$$

Note that we have implicitly assumed that no translational symmetry breaking occurs.

The total mean field Hamiltonian is a sum of the non-interacting Hamiltonian, $H$, of Sec. I A 1 and this mean field term:

$$\begin{aligned} \mathcal{H}_{MF} &= \mathcal{H} + \mathcal{H}_{C,MF} = \sum_{\boldsymbol{k}} c^\dagger_{i\boldsymbol{k}} \tilde{H}_{ij}(\boldsymbol{k}) c_{j\boldsymbol{k}} - \frac{1}{2\mathcal{V}} \sum_{\boldsymbol{k}} \left\langle \mathrm{tr}\left( c^\dagger_{\boldsymbol{k}} W c_{\boldsymbol{k}} \right) \right\rangle, \\ \tilde{H}_{ij}(\boldsymbol{k}) &= H_{ij}(\boldsymbol{k}) + W_{ij}. \end{aligned} \tag{S26}$$

We diagonlize $\tilde{H}_{ij}(\boldsymbol{k})$ with the matrices $U_{ai}(\boldsymbol{k})$:

$$\sum_{i,j} U^\dagger_{ai}(\boldsymbol{k}) H^{(MF)}_{ij}(\boldsymbol{k}) U_{jb}(\boldsymbol{k}) = \delta_{ab} \epsilon_a(\boldsymbol{k}). \tag{S27}$$

The matrix $W$ is obtained by substituting $c_{i\boldsymbol{k}} = \sum_a U_{ia}(\boldsymbol{k}) f_{a\boldsymbol{k}}$:

$$W_{ij} = -\frac{2E_C}{\mathcal{V}} \sum_{\boldsymbol{p}} \sum_{i,j} L_{ij} U^{\dagger}_{aj}(\boldsymbol{p}) \left\langle f^{\dagger}_{a\boldsymbol{p}} f_{b\boldsymbol{p}} \right\rangle U_{ib}(\boldsymbol{p}) = -\frac{2E_C}{\mathcal{V}} \sum_{\boldsymbol{p}} \sum_{i,j} L_{ij} U_{ia}(\boldsymbol{p}) n(\epsilon_a(\boldsymbol{p})) U^{\dagger}_{aj}(\boldsymbol{p}) \tag{S28}$$

where $n(\epsilon)$ is the Fermi distribution. Importantly, the modes $f_{a\boldsymbol{k}}$ and the corresponding unitary matrices are implicitly functions of $W_{ij}$. Self-consistency requires that for a given electron filling fraction, $n_f$, $W_{ij}$ is returned when the right-hand side of Eq. (S28) is calculated using the matrices which diagonalize $\tilde{H} = H + W$.

Numerically, we solve this iteratively for every filling $n_f$ under consideration. We start with $\tilde{H}^{(0)}(\boldsymbol{k}) = H(\boldsymbol{k})$. Diagonalizing, we obtain the energies $\epsilon^{(0)}(\boldsymbol{k})$ and matrices $U^{(0)}(\boldsymbol{k})$, allowing us to determine the chemical potential $\mu^{(0)}$ corresponding to the desired filling $n_f$. With this, we can then compute $W^{(0)}$. We subsequently define a new mean field Hamiltonian $\tilde{H}^{(1)}(\boldsymbol{k}) = H(\boldsymbol{k}) + W^{(0)}$. Again, we diagonalize to obtain new energies and rotation matrices, $\epsilon^{(1)}(\boldsymbol{k})$ and $U^{(1)}(\boldsymbol{k})$, allowing us to determine the chemical potential, $\mu^{(1)}$, corresponding to $n_f$. Inserting this data into the right-hand-side of Eq. (S28), we find a new matrix $W^{(1)}$. If $||W^{(1)} - W^{(0)}|| = 0$ up to some threshhold, we have found the mean field solution. If not, we repeat these steps until $||W^{(n)} - W^{(n-1)}|| = 0$ is satisfied.

In general more than one self-consistent solution $W$ exists. Further, some of these solutions result in ground states which spontaneously break some of the symmetries of the Hamiltonian. To take these into account, we alter the non-interaction Hamiltonian in the first step, $\tilde{H}^{(0)}(\boldsymbol{k}) = H(\boldsymbol{k}) + \delta H(\boldsymbol{k})$, where $\delta H(\boldsymbol{k})$ transforms non-trivially under one or more of the symmetries. While $\delta H(\boldsymbol{k})$ is not included in any subsequent step, initializing in this fashion allows the simulation to find symmetry-broken solutions should they exist.

More explicitly, this perturbation takes the form

$$\delta H(\boldsymbol{k}) = \sum_G \alpha_G \mathcal{O}_G(\boldsymbol{k}) \tag{S29}$$

where $G$ is a symmetry group element, $\alpha_G$ are small numbers relative to the flat bandwidth, and $\mathcal{O}_G(\boldsymbol{k})$ are the momentum components of an operator, or “order parameter,” which transforms nontrivially under the symmetry $G$ and trivially under all others. Ideally, we would like to restrict to the generators of the symmetry group: $G = C_2\mathcal{T}$, $C_3$, $M_{2y}$. However, it turns out that the best way to seed $C_2\mathcal{T}$ symmetry-breaking is through a joint $C_2\mathcal{T}$, $M_{2y}$ order parameter. Because the flat band wavefunctions are primarily located on the triangular lattice sites, we consider order parameters located on these sites as well. In particular, the ones we use are:

$$\begin{aligned}
\mathcal{O}_{C_2\mathcal{T}\cdot M_{2y}} &= \sum_{\boldsymbol{r}} \left(\tau^\dagger_{+,\boldsymbol{r}}\tau_{+,\boldsymbol{r}} - \tau^\dagger_{-,\boldsymbol{r}}\tau_{-,\boldsymbol{r}}\right) = \sum_{\boldsymbol{k}} \left(\tau^\dagger_{+,\boldsymbol{k}}\tau_{+,\boldsymbol{k}} - \tau^\dagger_{-,\boldsymbol{k}}\tau_{-,\boldsymbol{k}}\right), \\
\mathcal{O}_{C_3} &= \sum_{\boldsymbol{r}} \left(\tau^\dagger_{+,\boldsymbol{r}+\boldsymbol{a}_2} + \tau^\dagger_{+,\boldsymbol{r}-\boldsymbol{a}_1-\boldsymbol{a}_2} + \tau^\dagger_{+,\boldsymbol{r}+\boldsymbol{a}_1}\right)\tau_{-,\boldsymbol{r}} + h.c. \\
&= \sum_{\boldsymbol{k}} \left(\phi_{01} + \phi_{10} + \phi^*_{11}\right)\tau^\dagger_{+,\boldsymbol{k}}\tau_{-,\boldsymbol{k}} + h.c. \\
\mathcal{O}_{M_{2y}} &= \sum_{\boldsymbol{r}} i\left(\tau^\dagger_{+,\boldsymbol{r}+\boldsymbol{a}_1} + \omega\tau^\dagger_{+,\boldsymbol{r}+\boldsymbol{a}_2} + \omega^*\tau^\dagger_{+,\boldsymbol{r}-\boldsymbol{a}_1-\boldsymbol{a}_2}\right)\tau_{-,\boldsymbol{r}} + h.c. \\
&= \sum_{\boldsymbol{k}} i\left(\phi_{01} + \omega\phi_{10} + \omega^*\phi_{11}\right)\tau^\dagger_{+\boldsymbol{k}}\tau_{-\boldsymbol{k}} + h.c.
\end{aligned} \tag{S30}$$

where $\phi_{\ell m} = e^{-i(\ell k_1 + m k_2)}$.

### C. Results

At certain $n_f$, we find a number of different solutions to the mean field equation, and these can be classified by their symmetry-breaking behaviour. There are three general classes: those which preserve all the symmetries, those which break $C_3$ but preserve $C_2\mathcal{T}$ and $M_{2y}$ and those which break $C_2\mathcal{T}$ and $M_{2y}$ (we find that these two symmetries are always broken together). There are further distinctions within each class. For instance, for those solutions with $C_3$ symmetry breaking, two distinct symmetry breaking patterns are found – one of these is shown in Fig. 3 of the main text. Similarly, for the solutions which break $C_2\mathcal{T}$ and $M_{2y}$, there are two solutions which also break $C_3$ and one which does not.

The ground state at a filling $n_f$ is given by the solution with the lowest energy. In Fig. S8(a), we show the ground state energy $E_0$ per unit cell per flavour as a function of filling $n_f$ for these six solutions. For $n_f \lesssim -0.5$ and $n_f \gtrsim 0.5$, we see that the energies converge to a single value. Close to charge neutrality, the energies of the symmetry-broken states dip below that of the symmetric state in a non-trivial fashion. The difference between the ground state energy of the symmetry-broken solutions and the symmetric solution is displayed in Fig. S8(b).

We now discuss each of these solutions in turn.

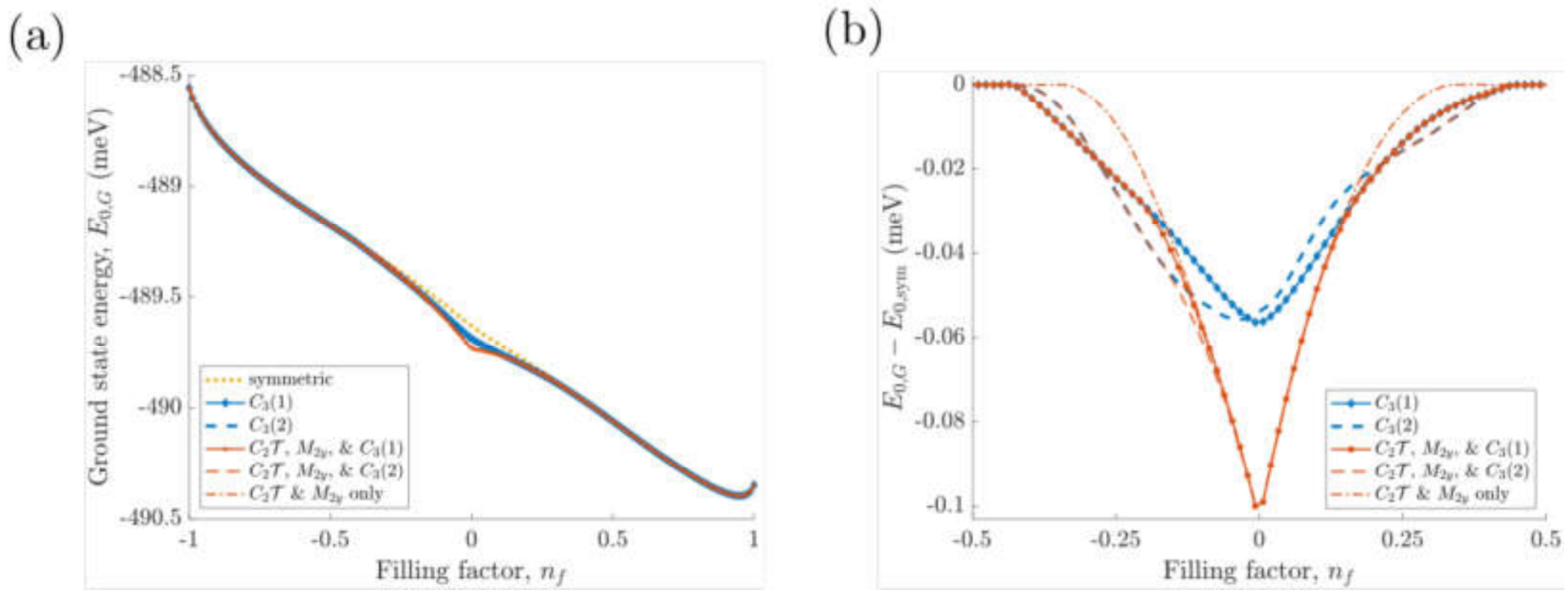

Figure S8. (a) Ground state energy per unit cell per flavour as a function of filling $n_f$ for the three general classes of solutions. The yellow dotted line plots the energy of the symmetric solution. The two blue lines display the energies of the two solution with $C_3$-breaking only, while the three red lines plot the energies of the $C_2\mathcal{T}$-broken solutions (including those which also break $C_3$). (b) The difference is ground state energies between the symmetry-broken solutions and the symmetric solution.

### 1. *Fully symmetric*

When we initialize with $\delta H = 0$, no symmetry-breaking can occur. We will refer to this solution as $\mathcal{S}$.sym. The resulting density of states as a function of filling and energy is shown in Fig. S9(a) while linecuts of the density of states at $n_f = 0$ and $n_f = 1$ are shown in (b). Both (a) and (b) indicate that the density of states changes very little as a function of filling.

### 2. $C_3$ *breaking only*

By initializing our Hamiltonian with $\delta H = \alpha_{C_3}\mathcal{O}_{C_3}$ ($\mathcal{O}_{C_3}$ is given in Eq. (S30)), depending on the sign of $\alpha_{C_3}$ we choose, two distinct $C_3$-breaking solutions are obtained close to charge neutrality. When $\alpha_{C_3} < 0$, we obtain the density of states shown in Fig. S10(a). Unlike the symmetric case, there is significant splitting between the peaks close to charge neutrality. This is further emphasized by Fig. S10(b), where linecuts of the DOS at $n_f = 0$ and $n_f = 1$ are shown. We refer to this set of solutions as $\mathcal{S}.C_3$(1). (We note that these figures have already been shown in Figs. 3(A)-(C) of the main text.)

Conversely, when $\alpha_{C_3} > 0$, a separate set of solutions, termed $\mathcal{S}.C_3$(2), is obtained. The result-ing DOS is shown in Fig. S10(e) and a linecut at $n_f = 0$ and $1$ is provided in (f). As with $\mathcal{S}.C_3$(1), both peak broadening and peak separation are observed in the region surrounding charge neutrality. Compared to Fig. S10(e), the splitting of the lower-band peak (left side of plot) is substantially less pronounced than it is for the $\mathcal{S}.C_3(1)$ solution in (a). This is also clear from the density linecuts in Figs. S10(b) and (f).

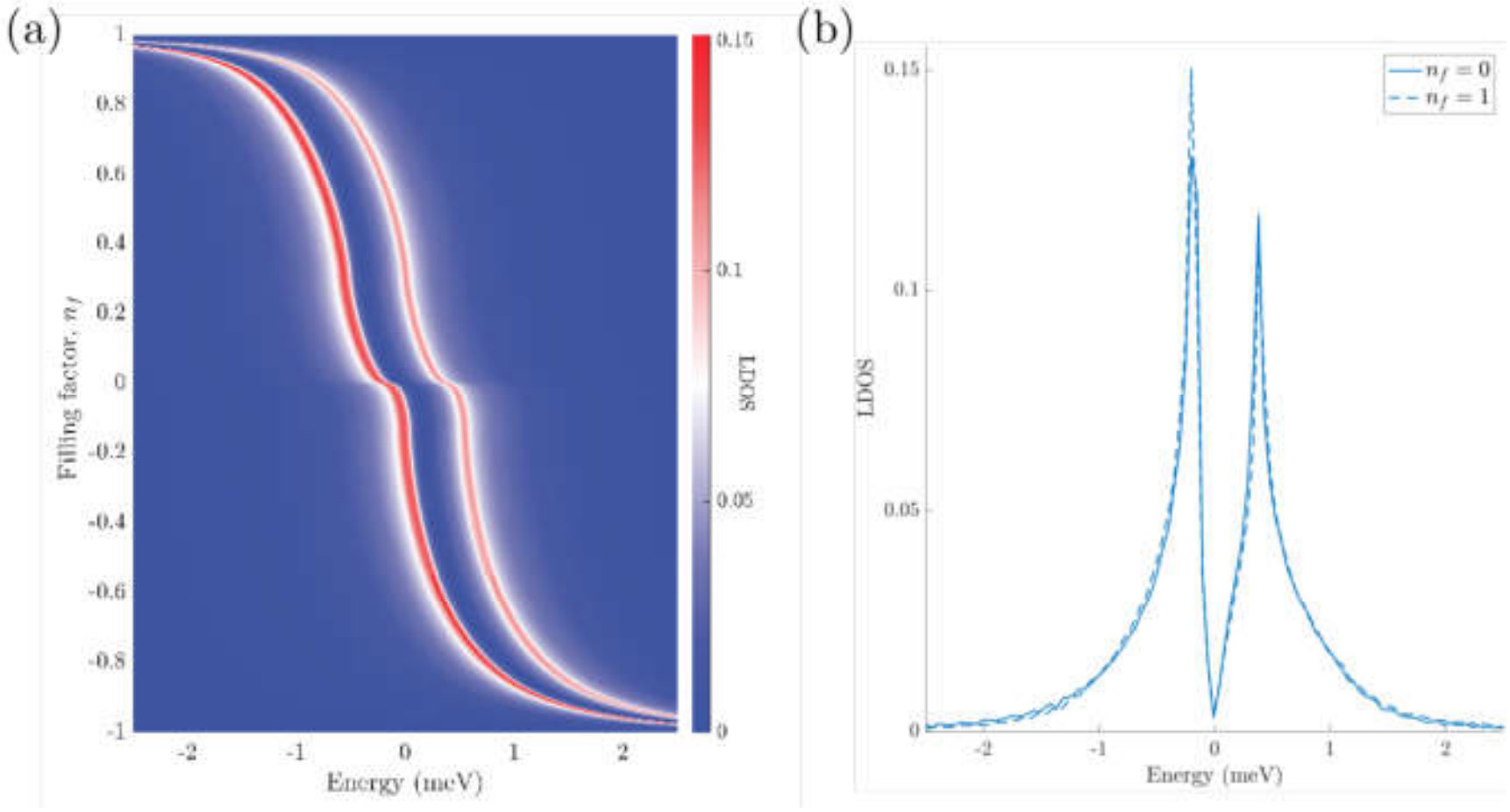

Figure S9. (a) Local density of states of the symmetry-preserving solution. (b) A linecut of the density at the charge neutrality point, $n_f = 0$, (solid line) and with a fully-filled flat band, $n_f = 1$ (dashed line). The linecut at $n_f = 1$ has been shifted so the Dirac points occur at zero energy.

We verify that the differences between Figs. S10(a) and (e) on the one hand and the DOS of the symmetric solution in Fig. S9(a) on the other can be accounted for by $C_3$ breaking by plotting the absolute values of the expectation values of the three order parameters in Eq. (S30). The results are shown in Figs. S10(c) and (g) for $\mathcal{S}.C_3$(1) and (2) respectively. In both, $\left|\left\langle \mathcal{O}_{C_2\mathcal{T}\cdot M_{2y}} \right\rangle\right|$ vanishes for all $n_f$, while $|\langle \mathcal{O}_{C_3} \rangle|$ is nonzero for $-0.5 \lesssim n_f \lesssim 0.5$. The conservation of $C_2\mathcal{T}$ and $M_{2y}$ and the breaking of $C_3$ has also been explicitly demonstrated by studying the symmetry transformation properties of the corresponding mean field Hamiltonians. We conclude that within the interval $-0.5 \lesssim n_f \lesssim 0.5$, only $C_3$ is broken and that outside of it, all symmetries are preserved.

These results are in conjunction with the observation that the density of states in Figs. S10(a) and (e) and Fig. S9(a) are identical when $n_f \lesssim -0.5$ or $n_f \gtrsim 0.5$. Analogously, in Fig. S8 the ground state energies of $\mathcal{S}.C_3$(1) and (2) converge to the same value as the symmetric solution for these fillings.

At charge neutrality, both $\mathcal{S}.C_3$(1) and $\mathcal{S}.C_3$(2) are lower in energy than the symmetric solution by $\sim 0.05$ meV. While Fig. S8 shows that when $n_f > 0$, $\mathcal{S}.C_3$(1) is preferred over $\mathcal{S}.C_3$(2) and vice versa for $n_f < 0$, the two solution are always very close in energy, with a maximal energy difference of $\sim 0.01$ meV. Given that strain is expected to be of order 0.03 meV (taking into account the scaling factor of 15), we should not attempt to distinguish the two based on energetics.

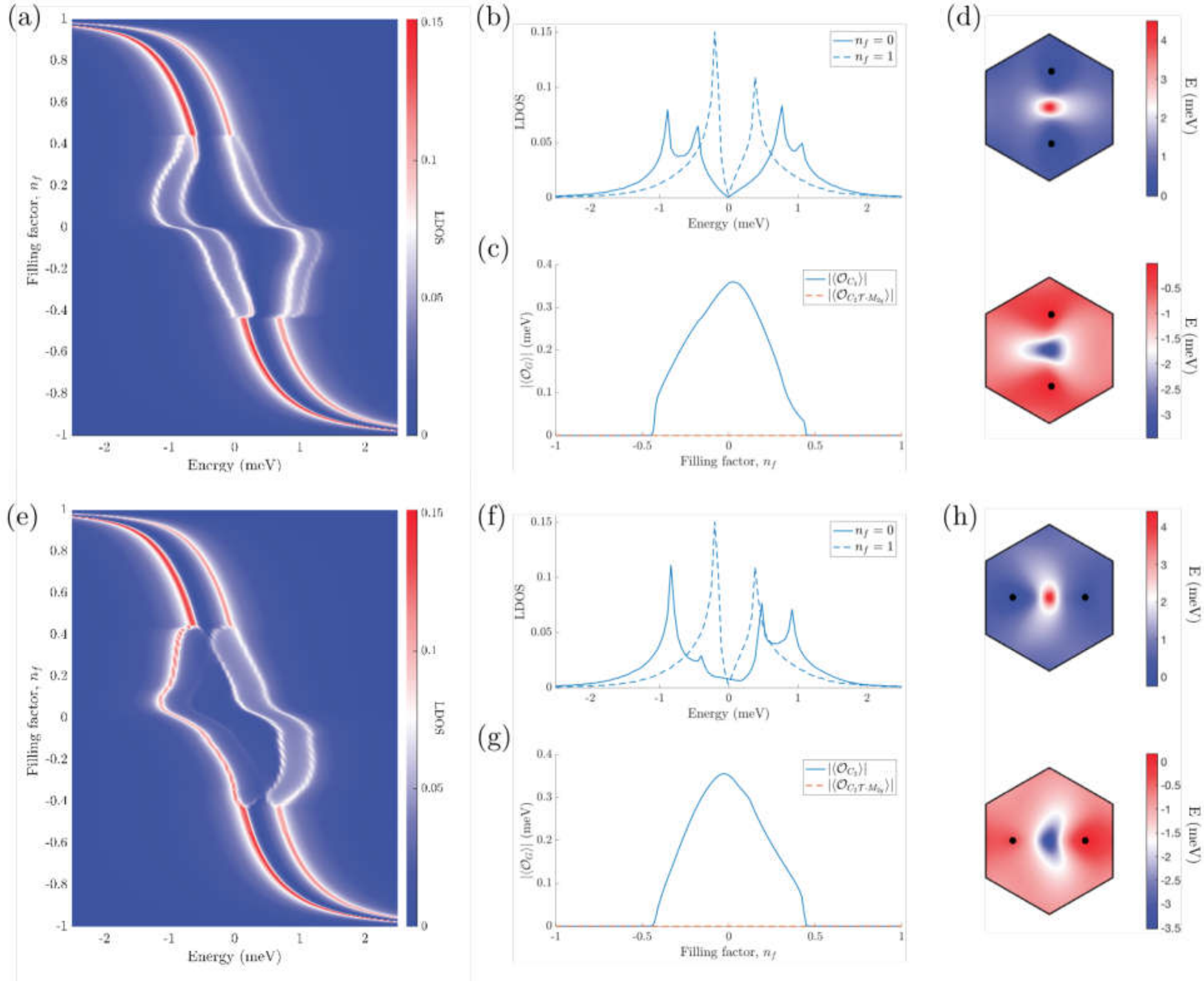

Figure S10. Plot of $C_3$-breaking solutions which preserve $M_{2y}$ and $C_2\mathcal{T}$. (a)-(d) correspond to $\mathcal{S}.C_3$(1), while (e)-(h) correspond to $\mathcal{S}.C_3$(2). (a),(b) Density of states as a function of filling and energy. (b),(f) Linecut at $n_f = 0$ (solid) and $n_f = 1$ (dashed). The linecut at $n_f = 1$ has been shifted so the Dirac points occur at zero energy. (c),(g) Magnitude of the $C_3$ and $C_2\mathcal{T} \cdot M_{2y}$ order parameter expectation values. (d),(h) Mean field band structure within the moiré BZ obtained at charge neutrality. The top plot corresponds to the flat band and the bottom plot to the lower flat band. The Dirac points are indicate by the black dots.

Since neither symmetry nor ground state energy can differentiate $\mathcal{S}.C_3$(1) and (2), we plot the band structures obtained at charge neutrality in Figs. S10(d) and (h). As mentioned above, the $C_2\mathcal{T}$ symmetry protects the Dirac cones, while $C_3$ protects their location. Accordingly, as indicated by the black dots, the Dirac cones remain but are no longer located at the $\boldsymbol{K}$ and $\boldsymbol{K}'$ points. Here, a clear difference between $\mathcal{S}.C_3$(1) and (2) is observed: The Dirac cones of $\mathcal{S}.C_3$(1) lie close to the $y$-axis and are interchanged under $M_{2y}$, but those of $\mathcal{S}.C_3$(2) are on the $x$-axis, and so are mapped to themselves under $M_{2y}$.

While both solutions are clearly $M_{2y}$ symmetric, the solution $\mathcal{S}.C_3$(2) cannot be continuously connected to the non-interacting band structure without breaking the $M_{2y}$ symmetry. This was our primary motivation for choosing to show the density of states given by $\mathcal{S}.C_3$(1) in the main text.

*a. Spatial map of local density of states* One way to probe $C_3$ breaking experimentally is to map out the spatial profile of the local density of states in real space at a given bias voltage by the STM tip. Theoretically, this local density of states $\rho(\boldsymbol{r};E)$ at location $\boldsymbol{r}$ and bias voltage $V_{\text{bias}} = E$ is encoded in the imaginary part of the real-space local Green's function, $G(E;\boldsymbol{r},\boldsymbol{r})$

$$\rho(\boldsymbol{r}) = -\frac{1}{\pi}\lim_{\eta\to 0^+} G(E+i\eta;\boldsymbol{r},\boldsymbol{r}). \tag{S31}$$

For a tight-binding model, such as the ten-band model used in this work, only the local density of states of a particular orbital on a given sublattice is well-defined. For a system with translational invariance with periodic boundary condition, the real-space local Green function for the ten-band model with mean-field-decoupled interaction is given by

$$G(E) = \frac{1}{\mathcal{V}}\sum_{\boldsymbol{k}}\left[E-\tilde{H}(\boldsymbol{k})\right]^{-1}, \tag{S32}$$

which is a $10\times 10$ matrix in the basis of the $10$ orbitals within a unit cell. Note that $G(E)$ is defined within a single unit cell and it will be the same in every unit cell across the whole system, due to translational invariance.

The local density of states $\rho_\alpha$ of a particular orbital $\alpha$ within a unit cell is defined as

$$\rho_\alpha(E) = -\frac{1}{\pi}\lim_{\eta\to 0^+} G_{\alpha\alpha}(E). \tag{S33}$$

Although the spectral weight of the flat bands is mainly located on the triangular sites, the $\rho_\alpha(E)$ with $\alpha$ corresponding to the three kagome sites does contain information on whether $C_3$ symmetry is present in the system. This is due to the fact that under $C_3$, the three orbitals on the kagome sites transform as $\kappa_1 \to \kappa_2$, $\kappa_2 \to \kappa_3$, and $\kappa_3 \to \kappa_1$, and so do the $\rho_\alpha(E)$ on these orbitals. Thus, the local density of states will be the same on the three types of kagome sites in our ten-band model when $C_3$ symmetry is present. However, when $C_3$ is broken but $M_{2y}$ is preserved, as in the case of Fig. 3 of the main text, $\rho_{\kappa_1}(E)$ can differ from $\rho_{\kappa_2}(E)$ and $\rho_{\kappa_3}(E)$, while the latter two are the same because of $M_{2y}$ symmetry. It is worth mentioning that, at least to some extent, the local density of states at the three kagome sites simulates the local density of states measured experimentally in between two moiré (triangular) sites, which are oriented along three directions. In Fig. 3F of the main text, we have plotted the $\rho_\alpha$ as widths of the bonds between triangular sites, which can be compared with the experimental map of the local density of states.

### 3. $C_2\mathcal{T}/M_{2y}$ *breaking*

We find three distinct solutions which break both the $C_2\mathcal{T}$ and $M_{2y}$ symmetries. Note that $C_2\mathcal{T}$-broken phases at neutrality were also found very recently in a continuum model with realistic Coulomb interaction (25). While the sign of $\alpha_{C_2\mathcal{T}\cdot M_{2y}}$ does not affect the state, different results are obtained depending on whether $C_3$ is allowed to break and, if so, the sign of $\alpha_{C_3}$. The density of states obtained by initializing with $\alpha_{C_3} < 0$, $\alpha_{C_3} > 0$, and $\alpha_{C_3} = 0$ are respectively shown in Figs. S11(a), (b), and (c). We refer to these solutions as $\mathcal{S}.C_2\mathcal{T}.C_3$(1), $\mathcal{S}.C_2\mathcal{T}.C_3$(2), and $\mathcal{S}.C_2\mathcal{T}$. (Since $M_{2y}$ is broken whenever $C_2\mathcal{T}$ is, for notational simplicity we do not include it in our naming convention.)

As observed in the previous section, away from charge neutrality, $n_f \gtrsim 0.5$ or $n_f \lesssim -0.5$, all three solutions converge to the symmetric solution $\mathcal{S}$.sym. Figures. S11(d) and (e) both demonstrate that for these fillings, the $C_3$ and $C_2\mathcal{T}\cdot M_{2y}$ order parameters vanish. Explicit study of the mean field Hamiltonians supports the conclusion that all solutions fully preserve the symmetries at these fillings.

We focus first on $\mathcal{S}.C_2\mathcal{T}.C_3$(1) and (2). Moving closer to charge neutrality, both Figs. S11(a) and (b) enter regions reminiscent of those found in Figs. S10(a) and (b). For $-0.5 \lesssim n_f \lesssim -0.2$ and $0.2 \lesssim n_f \lesssim 0.5$, only $C_3$ is broken, as indicated in Figs. S11(d) and (e). In this region, it turns out that $\mathcal{S}.C_2\mathcal{T}.C_3$(1) and (2) are in fact identical to $\mathcal{S}.C_2\mathcal{T}.C_3$(1) and (2) respectively. Around $n_f \sim \pm 0.25$, however, $\left|\left\langle \mathcal{O}_{C_2\mathcal{T}\cdot M_{2y}} \right\rangle\right|$ becomes nonzero [Fig. S11(d)]. With $C_2\mathcal{T}$ broken, the Dirac points cease to be protected, and Fig. S11(f) shows a gap opening at this point. Continuing to-wards charge neutrality, Figs. S11(d) and (e) plots the $C_3$ and $C_2\mathcal{T}$ order parameter expectation values decreasing and increasing, respectively, and we conclude that the $C_3$ and $C_2\mathcal{T}$ orders are competing. Eventually, very close to charge neutrality, $C_3$ symmetry is completely restored in both $\mathcal{S}.C_2\mathcal{T}.C_3$(1) and (2).

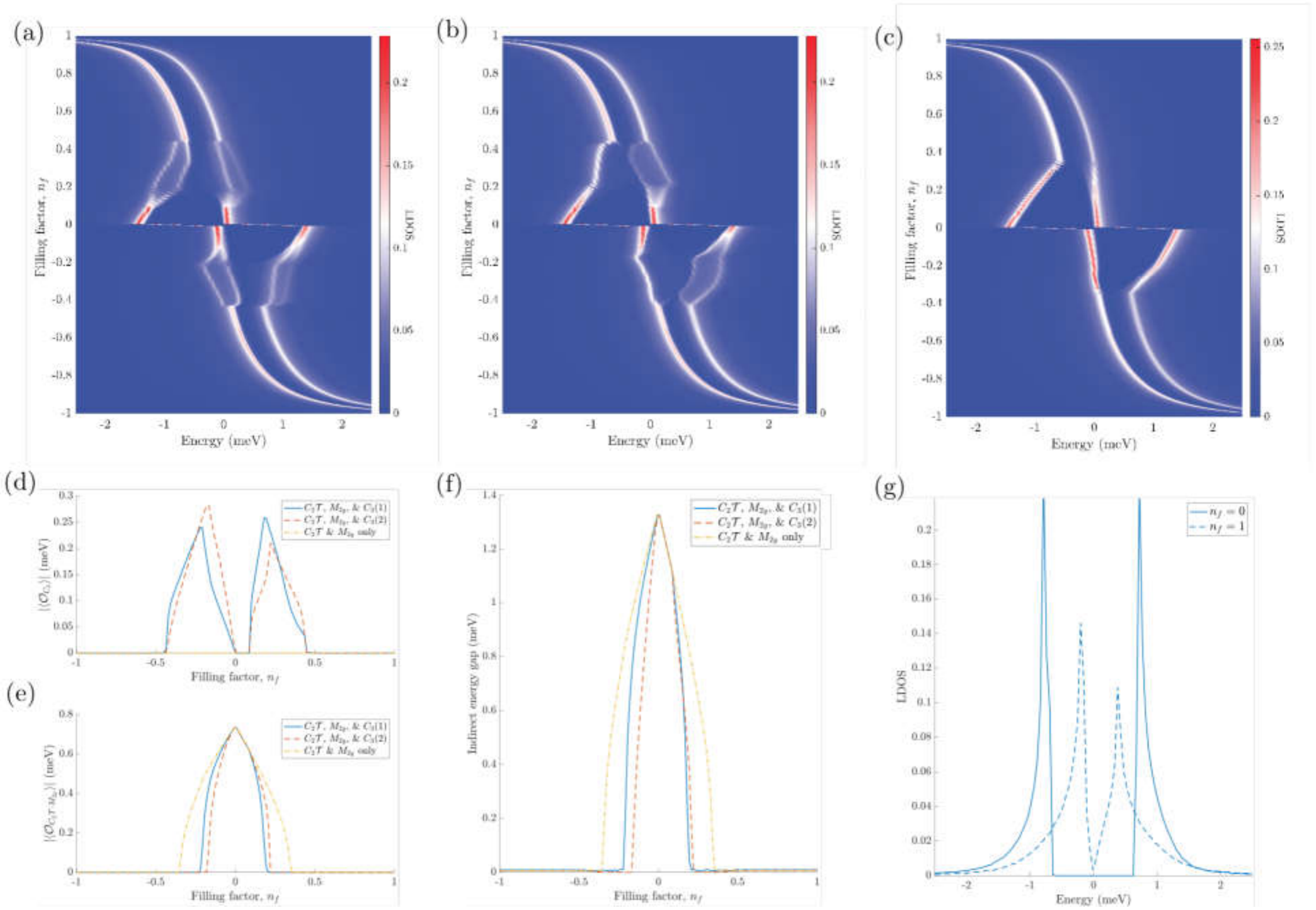

Figure S11. (a),(b),(c) Density of states for solutions $\mathcal{S}.C_2\mathcal{T}.C_3$(1), $\mathcal{S}.C_2\mathcal{T}.C_3$(2), and $\mathcal{S}.C_2\mathcal{T}$ respectively. (d) Magnitude of $C_3$ order parameter expectation values as a function of filling. (e) Magnitude of $C_2\mathcal{T} \cdot M_{2y}$ order parameter expectation values as a function of filling. (f) Gap as a function of filling. (g) Linecut at $n_f = 0$ and $n_f = 1$. For these fillings, the three solutions in (a)-(c) are identical. The linecut at $n_f = 1$ has been shifted so the Dirac points occur at zero energy.

Turning to the remaining solution, $\mathcal{S}.C_2\mathcal{T}$, the plot of the magnitude of the $C_3$ order parameter in Fig. S11(d) confirms that the $C_3$ symmetry is not broken for any filling fraction studied. Without having to compete with the $C_3$ order, the $C_2\mathcal{T}$ symmetry is able to break further from charge neutrality, close to $n_f \sim \pm 0.5$ as indicated in Fig. S11(e). Again, this coincides with the opening of the gap in (f).

Finally, at charge neutrality, all three solutions are identical and a gap is found. This is shown in Fig. S11(g).

### D. Discussion

Based on the ten-band model with local Coulomb interaction, and within a Hartree-Fock mean-field approximation, we have found various symmetry-broken solutions. Although the $C_2\mathcal{T}$-broken gapped solutions are lower in ground state energy per unit cell by about $0.05$ meV, the gapless $C_3$-broken solutions seem to be most consistent with experiment since no gap at charge neutrality is observed. The emergence of the gapless solution in experiment is plausible even when taking into account the scaling factor $\sim 15$ between our model and the physical system. With this factor, the energy difference translates into $\sim$ 7 K, which should be smaller than the error incurred by approximating the physical system with a simple ten-band model and a specific local interaction. Finally, in a real system the $C_3$ symmetry is presumably broken explicitly because of strain in the sample, which may make the $C_3$-broken solution more likely. On the other hand, breaking $C_2\mathcal{T}$ explicitly may be much harder than $C_3$. Suppose that the system exhibits long-range, $\mathcal{T}$-preserving disorder. In this case $C_2\mathcal{T}$ can only be explicitly broken if the disorder violates $C_2$. Crucially, $C_2$ map the valleys $\boldsymbol{K}_{\text{lbz}}$ and $\boldsymbol{K}'_{\text{lbz}}$ of the microscopic honeycomb system to one another, and, as emphasized above, these are separated by a very large momentum compared to the flat-band scales. Hence lifting $C_2$ symmetry requires atomic-scale deformations, whereas disorder on the moiré lattice scale would be sufficient to break $C_3$ symmetry.